\documentclass[12pt,aps,epsf]{emulateapj}

\usepackage{amsmath}
\shorttitle{Dynamical friction around SMBHs}
\shortauthors{Antonini and Merritt}
\def\beq{\begin{equation}}
\def\eeq{\end{equation}}
\def\fun#1#2{\lower3.6pt\vbox{\baselineskip0pt\lineskip.9pt
  \ialign{$\mathsurround=0pt#1\hfil##\hfil$\crcr#2\crcr\sim\crcr}}}
\def\lap{\mathrel{\mathpalette\fun <}}

\begin{document}
\title{Dynamical Friction around Supermassive Black Holes}
\author{Fabio Antonini} 
\email{antonini@astro.rit.edu}
\author{David Merritt}
\email{merritt@astro.rit.edu}
\affil{Department of Physics and Center for Computational Relativity and Gravitation, 
Rochester Institute of Technology, 85 Lomb Memorial
Drive, Rochester, NY 14623, USA}
\begin{abstract}
The density of stars in galactic bulges is often observed to be flat or slowly rising inside the influence radius of the supermassive black hole (SMBH).  Attributing the dynamical friction force to stars moving more slowly than the test body, as is commonly done, 
is likely to be a poor approximation in such a core since there are no stars moving more slowly than the local circular velocity.  We have tested this prediction using large-scale $N$-body experiments.  The rate of orbital decay never drops precisely to zero, because stars moving faster than the test body also contribute to the frictional force. When the contribution from the fast-moving stars is included in the expression for the dynamical friction force, and  the changes induced by the massive body on the stellar distribution are taken into account, Chandrasekhar's theory is found to reproduce the rate of orbital decay remarkably well. However, this rate is still substantially smaller than the rate predicted by Chandrasekhar's formula in its most widely-used forms, implying longer time scales for inspiral.
Motivated by recent observations that suggest a parsec-scale core around the Galactic center SMBH, we investigate the evolution of a population of stellar-mass black holes (BHs) as they spiral in to the center of the Galaxy.   After $\sim 10$ Gyr, we find that the density of BHs can remain substantially less than the density in stars at all radii; we conclude that it would be unjustified to assume that the spatial distribution of BHs at the Galactic center is well described by steady-state models.
One consequence is that rates of capture of BHs  by the SMBH at
the Galactic center (EMRIs) may be much lower than in standard
models. When capture occurs, inspiraling BHs often reach the
gravitational-radiation-dominated regime while on orbits that are still
highly eccentric;  even after the semi-major axis has decreased to
values small enough for detection by space-based interferometers,
eccentricities can be large enough that the efficient analysis of
gravitational wave signals would require the use of eccentric templates. 
We finally  study the orbital decay  of satellite galaxies into the central region of giant ellipticals and  discuss the
formation of multiple nuclei and multiplet of  black holes  in such systems.
\end{abstract}
\subjectheadings{ black hole physics-Galaxy:center-Galaxy:kinematics and dynamics-stellar dynamics}

\section{introduction}
Dynamical friction plays a central role in many astrophysical contexts. 
It drives the orbital inspiral and merger of satellite galaxies \citep[e.g.][]{MF:1980,IL:98,VDB:99} and
 the formation of massive black hole binaries \citep[e.g.][]{Q96, MM, MF04}, 
and it is the fundamental mechanism leading to mass segregation in dense stellar systems \citep[e.g.][]{BW:77,Frig:06,HA:06}.

Chandrasekhar formulated the principle of dynamical friction under the assumptions of an infinite, homogeneous and isotropic field of stars  (Chandrasekhar 1943). 
Despite these simplifications, his theory has been shown to work remarkably well 
in a wide variety of more general situations. 
Dynamical friction can be understood as the drag induced on a test particle by the overdensity (i.e., the gravitational wake) that is raised behind it by the deflection of stars \citep{DC:57,kln:72,MU:83}. 
The surprisingly good agreement between theory and numerical results may be
attributed to the fact that the  wake is a local structure, and over
small spatial scales, the stellar background appears nearly  homogeneous \citep{w:86}.
On the other hand, numerical studies have revealed a few, astrophysically important
contexts in which Chandrasekhar's theory appears to break down. 
These include the deceleration of a rotating stellar bar \citep{w:85},
inspiral in harmonic (constant-density) cores \citep{hg:98,go:06,read:06,in:09},
and the orbital evolution of a displaced supermassive black  hole \citep{GM:08}.

In this paper, we present  a comprehensive  study of dynamical friction in 
the nuclei of galaxies containing a dominant central point mass. 
In particular, we investigate the case of shallow density profiles around
supermassive black holes (SMBHs).  
Such nuclei appear to be common, and perhaps even generic.
For instance, the luminosity profiles of bright elliptical galaxies
always exhibit  flat central cores \citep{F:94,L:95}. 
Even the Milky Way, which was long believed to have a steeply-rising mass
density near Sgr A$^*$, is now believed to have a parsec-scale core
 \citep{BSE,D:09,B:10}.
Similar models may also be applicable to dark matter halos, if the central
point mass is identified with the stellar spheroid \citep{BS:01,BE:01,sgh:05}.

Theoretical treatments of dynamical friction make a surprising prediction about the 
frictional force in such systems.
Essentially all of the decelerating force is predicted to come from stars that are moving
more slowly than the test body.
But the phase-space density of a galaxy with a shallow density cusp around a
SMBH falls to zero at low energies: below a certain radius (roughly 1/2 the
core radius), there are no stars locally that move more slowly than the
circular velocity at that radius.
Chandrasekhar's formula, in its most widely-used form, 
would predict no frictional force.
In the case of an eccentric orbit that passes in and out of the core, 
the frictional force would be small near
periapsis, leading to a rapid increase in orbital eccentricity -- the opposite of
the usual assumption.
Our numerical experiments reveal that the frictional force does not drop
precisely to zero in such nuclei.
We show that the evolution can be well described
by a more general form formula that includes a contribution
to the force from stars moving faster than the test mass.
In this sense, our results affirm the correctness of Chandrasekhar's physical picture,
but only if the proper field-star velocity distribution is used (as opposed to, say, a Maxwellian),
and only if the usual simplifying assumptions that lead to a neglect of the contribution
of the fast stars to the frictional force are relaxed.

In \S 2 we review Chandrasekhar's derivation of the dynamical friction force
and highlight the approximations that lead to the neglect of the contribution
from the fast-moving stars.
We also briefly discuss alternative treatments of dynamical friction.
In \S 3 we  use Chandrasekhar's  formulae to integrate the equations of motion  of a massive body and follow its inspiral into the center of a model designed
to represent the Galactic center.
In \S 4 we use large-scale $N$-body simulations to  test  the  theory
in the case of inspiral of massive objects in  a nuclear star cluster with a flat density profile.
\S 5 investigates  the formation of the gravitational wake in the self consistent simulations.
Applications of our results to a variety of astrophysical problems are discussed in \S 6, and
\S 7 sums up.

\section{Dynamical friction}

The motivation for the $N$-body experiments described in this paper
is the existence of physically interesting models of galactic nuclei
in which the standard dynamical friction formula predicts little,
or  zero, frictional force.
We begin in this section by re-deriving the standard formula, noting the simplifying
approximations that are usually made.
We then present the more general form of Chandrasekhar's formula that includes
 contributions from field stars of all velocities, not just those that move
more slowly than the test body at infinity, and we evaluate
the expected contribution from the fast-moving stars in our models.
We also compute how the fast- and slow-moving stars contribute differently
to the steady-state density wake, using a technique first applied by Mulder (1983).
Finally we comment on perturbative approaches to computing dynamical
friction that relax the assumption of an infinite homogeneous medium.
The results obtained in this section constitute a set of baselines against which the
$N$-body results can be compared.

\subsection{Chandrasekhar's treatment}

\cite{CH:43} derived the coefficient of dynamical friction by summing the encounters
of a test body with passing stars, assuming that the unperturbed motion of the test
body was linear and unaccelerated, and that the field star distribution was 
infinite and homogeneous spatially and isotropic in velocity space.

The velocity change of a test body of mass $M$ in one encounter with a field star of mass $m\ll M$ is
\begin{equation}
\Delta v_{\parallel} = -2V{m\over M}{1\over 1+p^2/p_0^2}
\label{dvp}
\end{equation}
where $V$ is the relative velocity at infinity, $p$ is the impact parameter, and $p_0\equiv GM/V^2$.
The velocity change in equation~(\ref{dvp}) is parallel to the initial, relative
velocity $\boldsymbol{V}$ before the encounter.
In order to derive the coefficient of dynamical friction, one
sums the velocity changes in the direction of motion of the test mass, 
per unit interval of time, over all impact parameters
and over all values for the relative velocity at infinity.
The summation over impact parameters, at fixed $V$, is achieved by 
multiplying equation~(\ref{dvp}) by $2\pi pnVdp$, 
with $n$ the number density of field stars, and integrating $dp$:
\begin{equation}\label{eq:Dvp}
\overline{(\Delta v_{\parallel})} = -{2\pi G^2Mm n\over V^2}\ln\left(1+p^2_{max}/p_0^2\right).
\end{equation}
Under the assumption that $\Lambda\equiv p_\mathrm{max}/p_0\gg 1$, 
this can be written as
\begin{equation}
\overline{(\Delta v_{\parallel})} = -{4\pi G^2Mm n\over V^2}\left[
\ln\Lambda + \frac{1}{2}\frac{p_0^2}{p^2_{\rm{max}}} + ...\right].
\label{coef1}
\end{equation}
Terms beyond the first in brackets, the so-called ``non-dominant'' terms,
are usually neglected.

Returning to the more general form (\ref{eq:Dvp}),  
the dynamical friction coefficient is obtained by a second integration over
field star velocities $\boldsymbol{v}_\star$.
The relative velocity is $\boldsymbol{V}=\boldsymbol{v}-\boldsymbol{v}_\star$,
with $\boldsymbol{v}$ the velocity of the test star.
Since equation~(\ref{eq:Dvp}) gives the velocity change in the direction
of the initial relative motion, it must be multiplied by 
\beq
\frac{\boldsymbol{V}\cdot\boldsymbol{v}}{Vv} = 
\frac{v-v_{\star}}{V}
\eeq
to convert it into a velocity change in the direction of the test star's
motion, assumed here to be along the $x$ axis.
Let$f(\boldsymbol{v}_\star)\boldsymbol{dv}_\star$ be the number density of
field stars in velocity increment $\boldsymbol{v}_\star, \boldsymbol{v}_\star+\boldsymbol{dv}_\star$,
normalized to unit total number.
The dynamical friction coefficient is
\begin{eqnarray}\label{Equation:dfcoef}
\langle\Delta v_{\parallel}\rangle & = & \int f({\boldsymbol v}_\star)\, \overline{(\Delta v_{\parallel})}\,{v-{v_\star}_x\over V}\boldsymbol{dv}_\star \label{df} \\
&&=-2\pi G^2M \rho\int f({\boldsymbol v}_\star) {v-v_{\star,x}\over V^3} \ln\left(1+{p_{max}^2 V^4\over G^2 M^2}\right)\boldsymbol{dv}_\star \nonumber
\end{eqnarray}
where $\rho=m n$.

Henceforth we assume that the field star distribution is isotropic in velocity space.
Following \cite{CH:43}, we represent
the velocity-space volume element in terms of $v_\star$ and $V$ using
\beq
v-{v_\star}_x = {V^2+v^2-v_\star^2\over 2v}.\nonumber
\eeq
The result is
\begin{subequations}
\label{Equation:df2}

\begin{eqnarray}
\langle\Delta v_{\parallel}\rangle &=& 
-\frac{2\pi^2 G^2M\rho}{v^2}\int_0^{\infty} dv_\star\, v_\star\, f(v_\star)  {\cal H}\left(v,v_\star,p_\mathrm{max}\right),  \\
&& {\cal H}(v,v_\star,p_{max}) ={1\over 8v_\star}\int_{|v-v_\star|}^{v+v_\star} dV\left(1 + {v^2-v_\star^2\over V^2}\right)  ~~~~~~\\
&&\times \ln\left(1 + {p^2_{max}V^4\over  G^2M^2}\right)~. \nonumber
\label{Equation:df2b}
\end{eqnarray}
\end{subequations}
(The quantity $J$ defined in equation~(26) of Chandrasekhar (1943) is equal to $8v_\star {\cal H}$.)
The integral that defines ${\cal H}$ turns out to have an analytic solution;
the expression is complicated and we do not reproduce it here.
Chandrasekhar (1943) gave several approximate forms for ${\cal H}$ 
valid for $ p_{\rm max}/p_{\rm 0}  \gg 1$, e.g. his equation~(30):
\begin{equation}
{\cal H} \approx \left\{ \begin{array}{lllll}
	\ln {p_{max}\over GM} (v^2-v^2_\star) & \mbox{if $v>v_\star$,}\\ \\
	 {1\over 2} \ln \left(  4 {p_{max}\over GM~} v^2_\star\right)   -1 & \mbox{if $v=v_\star$,} \\ \\
 {\rm ln} \left( \frac{v_\star+v}{v_\star-v} \right)-2\frac{v}{v_\star} & \mbox{if $v<v_\star$}~.
	\end{array}
	\right.
	\label{Hap}
\end{equation}

In the standard approximation \citep[e.g.][]{RMJ:57},
 the non-dominant terms are set to zero, 
and the velocity dependence of the logarithmic term in the integrand
of equation~(\ref{Equation:df2}) is ignored.
Instead, one writes
\begin{equation}
\ln\left(1+{p^2_{max}V^4\over G^2M^2}\right) = 2\ln\Lambda  \equiv 
2\ln\left({p_{max}\over p_{min}}\right)
\end{equation}
 and the lower bound $p_\mathrm{min}$ 
is set to $GM/v_{\star,rms}^2$.
The weighting function ${\cal H}$ then takes on the simple form
\begin{equation}
{\cal H} = \left\{ \begin{array}{ll}
	\ln\Lambda & \mbox{if $v>v_\star$,} \\
	0 & \mbox{if $v<v_\star$}
	\end{array}
	\right.
	\label{eq:twoh}
\end{equation}
and the coefficient of dynamical friction is
\begin{equation}
\langle\Delta v_{\parallel}\rangle = -4\pi G^2 M\rho \times 4\pi \int_0^{v}dv_\star \left({v_\star\over v}\right)^2 f(v_\star).
\label{eq:dfh1}
\end{equation}
Equation (\ref{eq:dfh1}) reproduces the well-known result that only field stars with $v_\star<v$ contribute to the frictional force.

In this paper, we consider models for galactic nuclei in which
the number of stars moving more slowly than the test body can be
vanishingly small.
In such models, one expects that a significant fraction of the frictional
force might come from stars with $v_\star>v$.

The distribution of field-star velocities in our models
has the following form within the core:
\begin{equation}
f(v_\star) = \left\{ \begin{array}{ll}
	f_0\left(2v_c^2 - v_\star^2\right)^{\gamma-3/2} & \mbox{if $v_\star<2^{\frac12}v_c$,} \\
	0 & \mbox{if $v_\star>2^{\frac12}v_c$}
	\end{array}
	\right.
	\label{eq:fofv}
\end{equation}
where the normalizing constant
\beq
f_0 = \frac{\Gamma(\gamma+1)}{\Gamma(\gamma-\frac12)}
\frac{1}{2^\gamma\pi^{3/2}v_c^{2\gamma}}
\eeq
corresponds to unit total number.
This expression is equivalent to equation~(\ref{df}); it gives the local
distribution of velocities at a radius where the circular velocity
is $v_c=(GM_\bullet/r)^{1/2}$, assuming the density of field stars
follows $r^{-\gamma}$.
The phase space density is zero for $v_\star\ge v_\mathrm{esc}=2^{1/2}v_c$.

Of more interest here is the behavior of $f$ at small values of
$v_\star$, and when $\gamma<3/2$; for such values of $\gamma$ the phase
space density diverges at $v_\star=2^{1/2}v_c$.
As $\gamma\rightarrow 1/2$, the velocity distribution becomes progressively narrower,
and in the limit, $f(v_\star)$ is a delta-function at $v_\star=2^{1/2}v_c$;
in other words, all stars have zero energy.
This may be seen as a consequence of the well-known fact that 
$\rho\propto r^{-0.5}$  is  the shallowest power law density profile consistent with an isotropic velocity  distribution in a point-mass potential.

In the case of a test body moving in a circular orbit with $v=v_c$, the number
of field stars with $v_\star<v$ will drop as $\gamma$ approaches $1/2$,
and will equal zero in the limiting case $\gamma=1/2$.
The standard dynamical friction coefficient, equation~(\ref{eq:dfh1}), predicts
zero frictional force in this limit.

\begin{figure*}
\begin{center}
  \includegraphics[angle=-90.,width=4.5in]{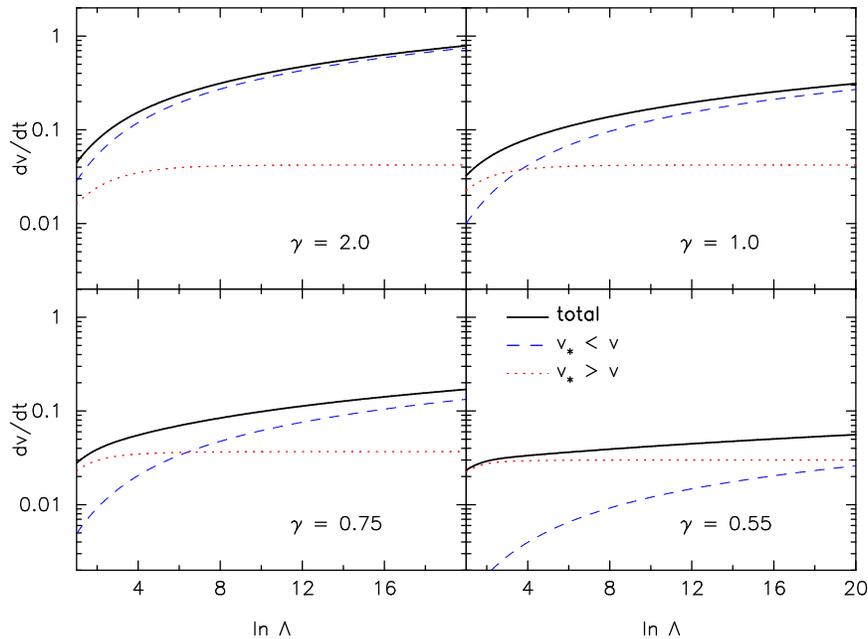}
  \caption{Contribution to the total dynamical friction force
from stars moving faster, or more slowly, at infinity than the
test body, assuming the velocity distribution of equation~(\ref{eq:fofv}).
The test body is assumed to be moving at the local circular velocity $v_c$.
In these plots, the configuration-space density $\rho$ remains fixed as
$\gamma$ is varied.
\label{fig:dv2}}
\end{center}
\end{figure*}

In this situation, it is clearly of interest to compute the contribution of the
fast-moving stars to the total frictional force.
We did this by evaluating ${\cal H}$ in its ``exact'' form, equation~(\ref{Equation:df2b}).
Figure~\ref{fig:dv2} shows the results.
In addition to $\gamma$, the results depend on the parameter
\beq
\ln\Lambda \equiv \ln\left(\frac{p_\mathrm{max}v_c^2}{GM_\bullet}\right)
\eeq
which plays the role of Coulomb logarithm.
We note the following results.
\begin{itemize}
\item For $\gamma\gtrsim 3/2$, the contribution to the frictional force from
the fast-moving stars is negligible, particularly when $\ln\Lambda$ is also large.
\item For $\gamma\lesssim 3/2$, the fast-moving stars contribute a progressively
larger fraction of the total frictional force, particularly when
$\ln\Lambda$ is small.
\item When $\gamma=0.55$, near the limiting value, the total frictional force
is small, and almost all of it comes from stars with $v_\star>v$.
\item Whereas the contribution to the force from the slow-moving stars depends strongly
on $\gamma$, the contribution from the fast-moving stars is almost independent
of $\gamma$.
\end{itemize}

According to equation~(\ref{Hap}), 
the contribution of the fast stars must tend to zero
as $\ln\Lambda$ is made sufficiently large.
This is consistent with Figure~\ref{fig:dv2}; however, for
$\gamma\approx 0.5$, the value of $\ln\Lambda$ required for
the slow stars to dominate is far greater than any physically 
reasonable value. 

\subsection{Mulder's treatment}

The foregoing treatment highlighted the contribution of the fast-moving stars,
$v_\star>v$, to the total  frictional force.
However it did not provide much insight into why the two
populations contribute in such a different way to the force.
Of course, the $N$-body experiments described in this paper include both
populations of stars.
In the simulations, the field stars quickly establish a nearly steady-state
distribution in a frame moving with the test mass -- a ``dynamical-friction wake''~ \citep{DC:57,kln:72,MU:83}. 
The over-density in the wake is responsible for the decelerating force
that acts on the test body.
A large fraction of the mass in the wake must be contributed
by the fast stars, particularly in the case that the fast stars dominate 
the density at large distances.
Why then do these stars contribute relatively little to the frictional force?

One way to address this question is via the technique of \citet{Mulder:83}.
Mulder computed the steady-state distributions of stars around a moving test mass,
making essentially the same assumptions as made by \citet{CH:43}.
He did this by invoking Jeans's theorem in a frame moving with the test mass,
and showing that an isotropic $f(v_\star)$ at infinity could be expressed
in terms of two of the integrals of motion in the Kepler problem.
This then allowed him to compute the steady-state density, in the moving
frame, at all locations around the test mass.
The dynamical friction force followed from a second integration of the density
over space;
Mulder showed that the results for the frictional force so obtained were consistent with Chandrasekhar's predictions,
if $p_\mathrm{max}$ were associated with the maximum dimension of the
spatial grid used to carry out the force integration.

\begin{figure*}
\begin{center}
  \includegraphics[angle=-90.,width=6.in]{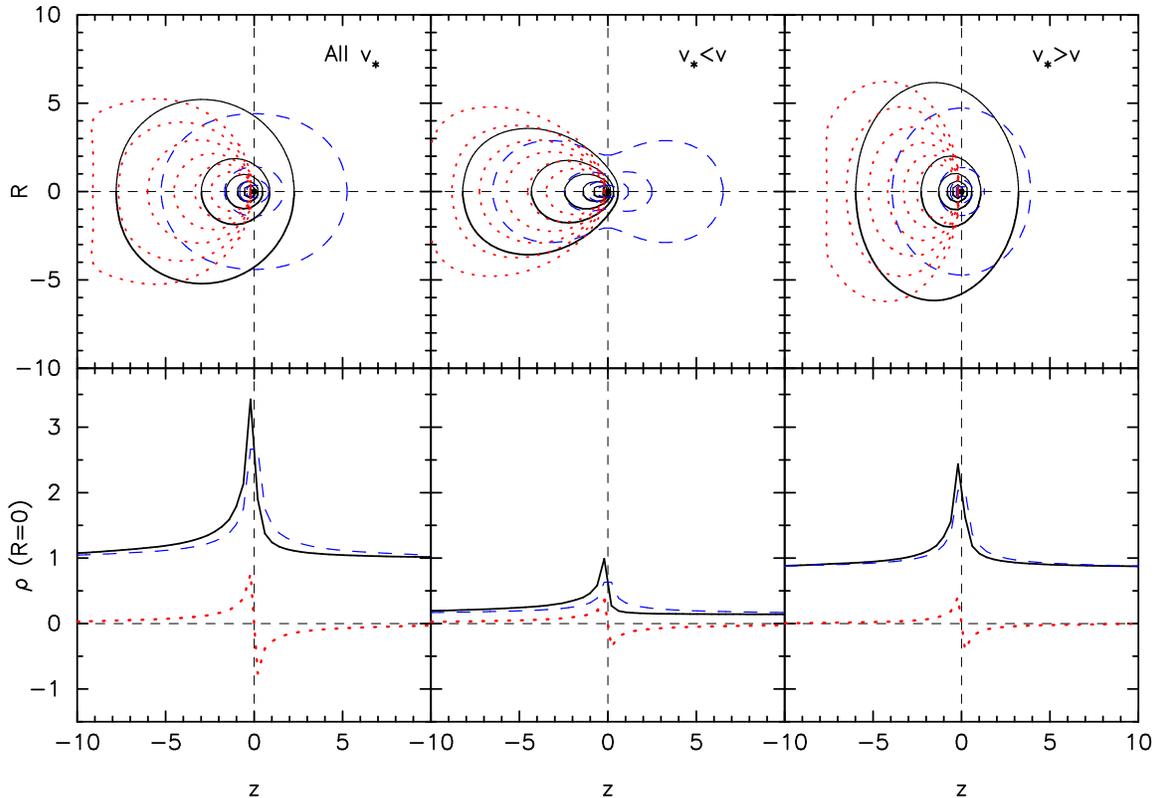}
  \caption{Dynamical friction wakes, computed via Mulder's
(1983) technique, assuming equation~(\ref{df}) with $\gamma=5/4$ for the velocity distribution at infinity; 
the test mass is located at the origin and is assumed to be moving at constant velocity $v=v_c$, as in Figure~1.
The top panels show contours of the density, in a plane that contains the
test body's velocity vector; the left panel shows the 
total density, the middle panel shows the density contributed by the
stars with $v_\star<v$ at infinity, and the right panel shows the contribution
from stars with $v_\star>v$ at infinity.
Black (solid) curves show the total response from the indicated stars;
blue (dashed) curves show the part of the response that is symmetric with
respect to $z$; red (dotted) curves show the asymmetric part (only on one
side), which is responsible for the frictional force.
The contours are spaced logarithmically in density and the contour spacing
is different in the three panels.
The lower panels show the density along the symmetry axis, i.e. along a line through
the test body in the direction of its motion.
Units are $G=M=v=1$.}
\label{fig:mulder}
\end{center}
\end{figure*}

Mulder's technique can be modified, to compute the separate contributions to 
the dynamical friction {\it wake} of the fast ($v_\star>v$) and slow ($v_\star<v$) stars;
here, as above, $v_\star$ refers to the field-star velocity at infinity.
The results are shown in Figure~\ref{fig:mulder}, for $\gamma=5/4$.
For this choice of $\gamma$, the fast stars dominate the total density at
infinity.
The density that they generate near the test body is also higher, 
everywhere along the symmetry axis, than the density due to the slow stars.
However the shapes of the two density wakes are very different:
in the case of the fast stars, the wake is elongated counter to the direction 
of the test body's motion, while in the direction parallel
to the motion, the change in density between the upstream and downstream
sides of the test mass is much less than in the case of the wake produced
by the slow stars.
These two differences are responsible for the small contribution of the
fast stars to the total frictional force (Figure~1), in spite of the higher density
of those stars at infinity and in the wake.

Comparison of the upstream and downstream densities in Figure~2 also suggests why the relative contribution of the fast stars to the frictional force drops off with increasing $\ln\Lambda$ in Chandrasekhar's treatment (Figure~1).
At large distances from the test body, the wake produced by the fast stars
is nearly symmetric; the greatest asymmetry is in the region near the test mass.
The wake generated by the slow stars, on the other hand, maintains its asymmetry much farther from the test body.
Roughly speaking, the density far from the origin in Figure 2 is produced by stars with large
impact parameters, and so increasing $p_\mathrm{max}$ in Chandrasekhar's
treatment corresponds to more heavily weighting the contribution from the slow-moving
stars.

\subsection{Perturbative treatments; inhomogeneous systems}

In treatments like Chandrasekhar's and Mulder's, 
the unperturbed trajectories consist of straight lines.
In reality, both test and field stars follow non-rectilinear orbits 
about the center of the galaxy.
Chandrasekhar's theory might be expected to give approximately correct
results even in this case, 
as long as $p_\mathrm{max}\gg p_\mathrm{min}$, 
since over many decades in scale the orbits of
the field stars will appear nearly rectilinear as seen by the test body.
But given certain assumptions, perturbation theory can be used
to more correctly compute the response of the orbits in a galaxy to the presence of
a perturbing potential \citep{LBK1972,TW1984,RT:96}.
One finds that the net torque on the test mass is due to orbits near
resonance, i.e. orbits for which the frequencies associated with the radial
and angular motions satisfy a relation $l_1\omega_r + l_2\omega_\theta - l_3\Omega_t=0$
where the $l_i$ are integers and $\Omega_t$ is the frequency of rotation of the
test mass (assumed to be on a circular orbit).
The acceleration induced by the resonant orbits depends on
how quickly the orbit of the test mass is evolving;
if orbital decay is very slow, the influence of a single resonance can build up,
invalidating the perturbative assumption, while if it is too fast, the assumption
of near-stationarity is violated.
Furthermore, in a real galaxy (or $N$-body system) the frequency spectrum of the perturbing potential is not made up of sharp lines, but rather is broadened by the time 
dependence of the  decaying orbit and by the finite age of the galaxy.

Due to the computational complexity involved, 
applications of this approach have so far been limited to bodies following circular orbits in
simple (Plummer, scale-free) galaxy models, and the results
have mostly been interpreted as corrections to the predictions of Mulder and Chandrasekhar.
For instance, \citet{w:86} emphasized the similarity in the structure
of the wake as computed via the perturbation formulae and via Mulder's
approach.
The main element that the perturbative treatments add is a quantitative estimate of the Coulomb logarithm.
Not surprisingly, none of these studies has attempted to relate the frictional force separately
to the ``fast'' and ``slow'' stars as they appear in Chandrasekhar's treatment;
doing so would be an ill-defined problem since all stars are included, self-consistently,
in the perturbative treatments.
Nevertheless, as far as we can tell, comparisons with Chandrasekhar's theory are
always made via equation~(\ref{eq:dfh1}), which ignores the fast-moving stars.

A potentially important application of the perturbative methods is to cases where the assumption of locality is violated.
For instance,  a satellite that orbits just outside a galaxy,
where the local density is zero, would experience no frictional force if
the local properties of the background were assumed to hold everywhere;
in reality it feels a force due to polarization of the orbits inside the galaxy
\citep{PP1985}.
The models considered in this paper constitute a second case
where the assumption of locality may be inappropriate, since some of the frictional
force acting on a test mass orbiting in the core will come from stars outside the core,
where $f(\boldsymbol{v})$ has a different functional form, including (for
instance) some slow-moving stars.
In lieu of such a calculation (and in view of the difficulties associated with
interpreting the results; e.g. \citet{Weinberg2004}), an $N$-body treatment seems
a logical first step.
As we will see, Chandrasekhar's formula, in its more general form,
turns out to reproduce the $N$-body results quite well.

\section{Orbital Evolution based on Chandrasekhar's Formulae}

We are interested in the orbital evolution of  a massive body
as it spirals in toward the center of a galaxy that contains a 
supermassive black hole (SMBH).
In subsequent sections, we present results from large-scale, direct-summation
 $N$-body simulations.
As a basis for comparison, we present in this section the predictions 
of Chandrasekhar's approximate formula.
We represent the stars via a smooth, fixed potential and integrate the equations 
of motion of the massive body in the fixed analytic potential including a term
that represents the non-conservative contribution of dynamical friction.

We base our model for the stellar density on the observed distribution of old stars
at the Galactic center (GC).
Number counts \citep{BSE,D:09,B:10}
are consistent with a density that follows a broken power-law:
\begin{equation}
\label{den}
\rho(r)=\rho_0 \left(  \frac{r}{r_0} \right)^{-\gamma}
 \left[
1+\left( \frac{r}{r_0} \right)^{\alpha}\right]^{(\gamma-\gamma_{e})/{\alpha}}~,
\end{equation}
where $\alpha$ is a parameter that defines the  transition strength between inner and
outer power laws  and $r_0$ is the scale radius.
 Following \citet{M:10}, we adopt  $r_0=0.3$pc , $\alpha=4$ and $\gamma_{e}=1.8$
as fiducial values.
The central slope $\gamma$ was left as a free parameter.
The normalizing factor $\rho_0$ 
was chosen in such a way that for each value of  $\gamma$, the corresponding 
density profile reproduces the coreless density model:
\begin{equation}
\label{fidmod}
\rho(r)=1.5\times 10^5 \left( \frac{r}{1{\rm pc}} \right)^{-1.8}{ M_{\odot} \mathrm{pc}^{-3}}
\end{equation}
outside the core.
This choice of normalizing constant
gives  a mass density at $1{\rm pc}$ similar to what various authors 
have inferred \citep[e.g.][]{okf} and implies a total mass in stars  within this 
radius of $\sim 1.6 \times 10^6 { M_{\odot} \mathrm{pc}^{-3}}$.

Assuming equal-mass stars  of mass $m$ and an isotropic velocity distribution, the local
two-body relaxation time is defined as \citep{Spitzer:87} :
\begin{equation}
\label{relax}
t_{\rm r}=\frac{0.33 \sigma^3}{\rho m G^2 {\rm ln} \Lambda }~,
\end{equation}
where $ {\rm ln} \Lambda$ is the Coulomb logarithm and
$\sigma$ is the isotropic velocity dispersion; the latter can be computed
from Jeans's equation,
\begin{equation}
\rho(r)\sigma(r)^2 = G\int_r^{\infty} dr' r'^{-2} \left[M_{\bullet}+M_\star(<r')\right]\rho(r').
\label{jeans}
\end{equation}
Here $M_{\bullet}$ is the mass of the central SMBH that we take to be $4 \times 10^6  { M_{\odot}}$ 
\citep{Ghez:08,Gillessen:09} and $M_\star(<r)$ is the total mass in stars within $r$.
The  total stellar mass  contained within the SMBH influence radius
($r_{\rm bh}\approx 2.5$pc) is $M_\star(<r_{\rm bh}) \approx 10^7 {M_{\odot}}$;
assuming solar-mass stars,
the two-body relaxation time at $r_\mathrm{bh}$ is 
$t_{\rm r}(r_{\rm bh})\approx 2\times10^{10}{\rm yr}$.

\subsection{Circular Orbits} \label{co}

\begin{figure*}
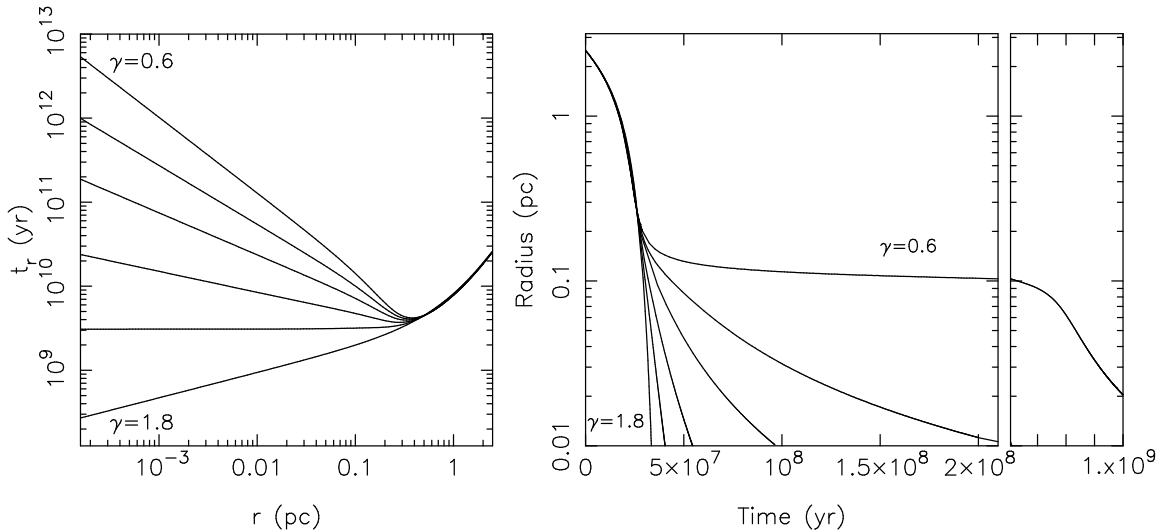

\begin{center}
$\begin{array}{cc}
\includegraphics[angle=0,width=2.56in]{fig3a.eps} &  \includegraphics[angle=0,width=3.38in]{fig3b.eps}
                    \end{array}$
  \caption{ \label{fig:dfcirc}
  Left panel: relaxation time $t_{\rm r}$ versus radius for  models based on the density
law of equation (\ref{den}).
  Right panel: Orbital decay of a $2 \times 10^3{M_{\odot}}$ massive body starting from a radius of $2.5 {\rm pc}$.  
  Here we used $\ln \Lambda=7$.
  In both panels, various values of the inner density slope $\gamma$ were considered: $(0.6,0.8,1,1.25,1.5,1.8)$.
\label{massmodel2}}
\end{center}
\end{figure*}

\begin{figure}
\begin{center}
  \includegraphics[angle=0,width=2.5in]{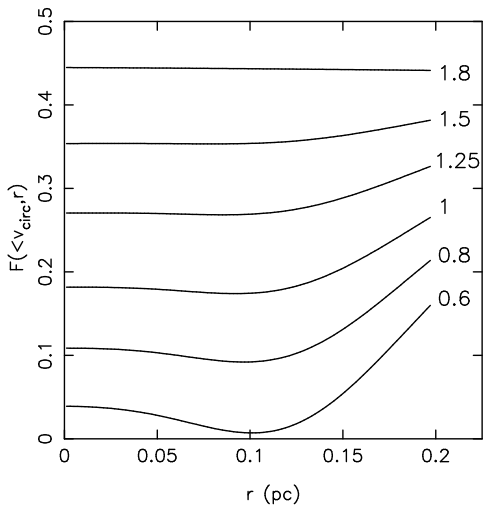}
  \caption{Fraction of stars $F(<v_{\rm circ},r)$ moving more slowly than the local circular velocity as a function 
  of radius for $\gamma=(0.6,0.8,1,1.25,1.5,1.8)$.
  When $\gamma=0.6$,  $F$ is close to  zero for $r \approx 0.1{\rm pc}$ .
  Hence, the frictional force acting  on a massive particle which moves on a circular orbit
   drops essentially to zero at this radius.
\label{NvsR}}
\end{center}
\end{figure}

The frictional acceleration  on a point particle of mass $M$
and velocity $\boldsymbol{v}$ is \citep{CH:43}
\begin{equation}
\boldsymbol{f_{\rm fr}} = -\frac{4\pi G^2  M \rho(r)F(<v,r)\ln\Lambda} {v^3} \boldsymbol{v},
\label{dfa}
\end{equation}
where $F(<v,r)$ is the fraction of stars at  $r$ that are moving more slowly than $v$.
This is the standard expression, derived by ignoring the velocity dependence of ln$\Lambda$ when integrating over the field-star velocity distribution and setting the non-dominant
terms to zero.   
As a result of these approximations, the frictional force 
is produced only by field stars with velocities less than $v$.
Although equation~(\ref{dfa})  was  derived under the assumptions of an infinite and 
homogeneous background of stars, it has been shown to work
reasonably well even for more general stellar distributions \citep{w:83,LT:83,w:86,cmv:97,MM:06,J:10}.

For a massive particle initially located at
 $r_{\rm bh}$ on a circular orbit, the inspiral time in the 
power-law density profile of equation ~(\ref{den}) with $\gamma=1.8$ (i.e., the coreless model)   is 
\begin{eqnarray}
 t_{\rm fr} \approx 6 \times 10^7 \mathrm{yr} \left(  \frac{r}{2.5 {\rm pc}}  \right)^2 
 \left( \frac{\sigma}{100 {\rm km s^{-1}}}  \right)  \nonumber \\
\times  \left( \frac{1 \times10^3 {M_{\odot}}} {M} \right) \left( \frac{7}{{\rm ln}\Lambda} \right) 
\end{eqnarray}
 independent of the mass of the field stars  if ${M}>>m$.

Figure~\ref{fig:dfcirc} plots the relaxation time as a function of radius for the same model,   assuming 
$\ln \Lambda=15$, $m={M_{\odot}}$ and adopting different values for
the inner density slope $\gamma$.
It turns out that the isotropic distribution function corresponding to 
the adopted density law (\ref{den})  becomes negative at certain energies
for $\gamma \lesssim 0.6$.
For this reason, we consider in the following only models with $\gamma \ge 0.6$.

Figure~\ref{fig:dfcirc} also shows the evolution of a $2\times 10^3{M_{\odot}}$ black hole on a
circular orbit starting from a galactocentric distance of $2.5~$pc and using $\ln \Lambda=7$. 
The orbit was  numerically integrated by solving 
the system of first-order differential equations
\begin{equation}
 \dot{\boldsymbol{r}}= \boldsymbol{v} , ~~ \dot{\boldsymbol{v}}=- \boldsymbol{\nabla} \phi +\boldsymbol{f}_{\rm fr} 
\label{em}
\end{equation}
with $\phi(r)$  the total gravitational potential  produced by the 
stars and the SMBH:
 \begin{eqnarray}
\phi(r) = -\frac{GM_{\bullet}}{r} + \phi_{\star}(r) =
-\frac{GM_{\bullet}}{r}  \nonumber \\
+ 4\pi G  \left[
\frac{1}{r}\int_0^{r} dr' r'^2 {\rm \rho(r')}
+ \int_r^{\infty} dr' r' {\rm \rho(r')}
\right]~.
\label{FS}
\end{eqnarray} 
The numerical integration was performed using a 7/8 order Runge-Kutta algorithm with a variable time-step \citep{F:68}
 in order to keep the relative error
per step in energy, in the absence of dynamical friction,
 less than a specified value ($10^{-8}$). 
 When  dynamical friction was included,  we checked the integration accuracy
 through the quantity  $E+E_{\rm df}$  with $E$ the energy per unit mass and 
 $E_{\rm df}$ the work done by dynamical friction along the trajectory.
The accuracy in this case was  
of the same order of that found in integrations without dynamical friction.
The function  $F(<v,r)$  was evaluated using the expression \citep{Szell:05}:
\begin{eqnarray}\label{fvr}
F(<v,r) = 1 - {1\over \rho}\int_{0}^E d\phi' {d\rho\over d\phi'} \nonumber \\
\times \left\{ 1 + {2\over\pi} \left[{v/\sqrt{2}\over\sqrt{\phi'-E}} - \tan^{-1}\left({v/\sqrt{2}\over\sqrt{\phi'-E}}\right)\right]\right\},
\end{eqnarray}
where $E=\frac{1}{2}v^2+\phi(r)$.

At all radii,   the relaxation time is  much longer than the 
time required for the massive particle to reach the core.
What happens next depends on $\gamma$: 
the orbital decay can essentially stall when $\gamma$ is small (i.e., $\sim 0.6$),
or continue rapidly if $\gamma$ is larger.

The explanation of this behavior can be found  in Figure~\ref{NvsR}
which plots  the fraction of stars moving 
more slowly than the local circular velocity $v_{\rm circ}(r)$  as a 
function of radius, for various values of $\gamma$. 
When $\gamma=0.6$,
$F(<v_{\rm circ},r)$ approaches  zero at $r_{\rm st} \sim 0.1{\rm pc}$ 
and consequently the dynamical friction force 
 drops drastically at this radius  (see equation [\ref{dfa}]). 
The stalling observed in the orbital evolution for this value of $\gamma$
 is therefore a consequence of the lack
of slowly-moving stars in the core.
However, the inspiral always continues into the very center
since $F(<v_{\rm circ},r)>0$ everywhere. 

 For $\gamma \geq 0.6$,   the  time required for dynamical friction  to bring
 a $10^3{M_{\odot}}$ black hole into the center, starting from a
 galactocentric distance  of a few parsecs,    is shorter than the two-body
  relaxation time  evaluated at the SMBH influence  radius $t_{\rm r}(r_{\rm bh})$.  
On the other hand, 
the dynamical friction force  
decreases with the mass of the inspiraling object,  and  for  
$M \lesssim 10^2 {M_{\odot}}$ 
 the infall timescale  can significantly exceed a Hubble time.
\citet{MS:06}  found  that $t_{\rm r}(r_{\rm bh})$ is also approximately the  
timescale over which  gravitational encounters change an initial density profile
into the Bahcall-Wolf   form, i.e., $\rho \propto r^{-1.75} $. 
We conclude that  
for a black hole of mass $M \ge 10^3 {M_{\odot}}$,
inspiral will occur in a mass profile that is almost independent of time.
However, for  $\gamma \sim 0.6$, the time required 
to reach a distance $\sim 0.01 {\rm pc}$, is still  comparable with the local relaxation time.
This will result in a substantial evolution of the stellar background  during the orbital decay. 

\subsection{Eccentric Orbits} \label{ecc}

\begin{figure*}
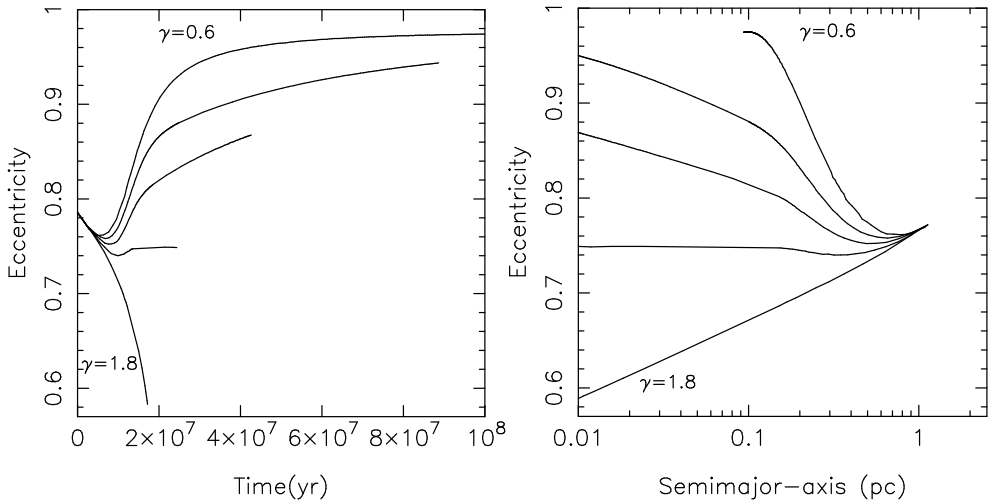

\begin{center}
$\begin{array}{cc}  
  \includegraphics[angle=0,width=2.62in]{fig5a.eps}   \includegraphics[angle=0,width=2.52in]{fig5b.eps}
                    \end{array}$           
  \caption{
  Left panel shows the time dependence of the  orbital eccentricity of a  $M = 2\times 10^3 {M_{\odot}}$
  black hole. In the right panel the orbital evolution is shown in the eccentricity-semimajor axis plane.
    The  inner cusp slopes are  $\gamma=(0.6,1,1.25,1.5,1.8)$.
     Initial   apoapsis and periapsis distances were  $2.5$ and $0.35$pc respectively
   and initial  semi-major axis was $a=1.4$pc. 
  The integrations terminated either  when the semi-major axis of the black hole was $0.01$pc
  or at $10^8$yr for  $\gamma=0.6$.
  }
\label{ea}
\end{center}
\end{figure*}

In the case of an isotropic distribution function $f(E)$  
describing a power law density profile
around a SMBH,  if the gravitational potential  produced by the stars is  ignored 
(i.e., $E<<-GM_\bullet/r_{\rm bh}$), then
\begin{equation}\label{dff} 
f(E)=\frac{3-\gamma}{8}\sqrt{\frac{2}{\pi^5}} \frac{\Gamma(\gamma+1)}{\Gamma(\gamma- 1/2)}  \times 
 \frac{M_\bullet}{m}  \frac{\phi_0^{3/2}}{\left(G M_\bullet \right)^3}\left( \frac{ |E|}{\phi_0} \right)^{\gamma-3/2},~~~~
\end{equation} 
with $\phi_0=GM_\bullet /r_{\rm bh}$ \citep{Merritt:12}.
For $\gamma \leq 0.5$, $f(E)$ is undefined and so $\gamma \approx 0.5$  is  the shallowest density profile consistent with an isotropic velocity distribution around a SMBH.
In the case   $\gamma=1.5$,  equation  (\ref{dff}) shows that
 the distribution function is a constant ($f(E) \equiv  f_0$).
 If one writes
\begin{equation}
\rho(r)F(<v,r) = \rho(r)\times {1 \over \rho(r)}  4 \pi \int_0^v dv_\star  v_\star ^2 f_0 = {4\over 3} \pi~f_0 v^3 
\end{equation}
it can be immediately seen that the product $\rho(r)F(<v,r)$ in equation  (\ref{dfa})  will be a function  of $v$ only (e.g., Just et al.~2011).
Under these circumstances,  the coefficient of dynamical friction
 will have only a weak  dependence  on radius  through  the Coulomb logarithm.
It can be shown that, in this case, the eccentricity of a massive body will remain unchanged during its  motion, while dynamical friction
 will either circularize the  orbit for $\gamma >1.5$ 
or make it  more eccentric for  $\gamma <1.5$  \citep{Q96, GQ03}.  

To evaluate the eccentricity evolution of a massive particle in response
to Chandrasekhar's dynamical friction formula, a numerical treatment is necessary.
We therefore carried out numerical integrations of  the set of differential equations
(\ref{em})  as described above, adopting as before equations~(\ref{den}) 
and~(\ref{FS}) for the (fixed) stellar potential.

Figure \ref{ea} shows the results for $M = 2\times 10^3 {M_{\odot}}$.
The massive particle was initially placed at $r=2.5$pc with a  tangential velocity of $\sim~0.36v_{\rm circ}$.
With this initial configuration the body    penetrates the  inner  core after few obits.
Different values of the internal slope $\gamma$, ranging from $1.8$ to $0.6$ were adopted. 
As a proxy for the instantaneous orbital elements, we computed over each radial period the 
largest and the smallest distance from the origin (i.e. the SMBH)  and defined these
as the apoapsis $r_{\rm ap}$ and periapsis $r_{\rm per}$ respectively.
The eccentricity and semi-major axis were then computed using the Keplerian
expressions
\begin{equation}
e=  \frac{r_{\rm ap}-r_{\rm per}}{r_{\rm ap}+r_{\rm per}}, \ \ \ \ 
a=\frac{r_{\rm ap}}{1+e} .
\end{equation}
The figure reveals a complex behavior of eccentricity on time. 
For $\gamma \le 1.5$ we distinguish three regimes.
In phase I,  the eccentricity decreases (even for $\gamma\ge 1.5$).
The duration of this phase is  shorter for shallower profiles. 
After reaching a  minimum,  the eccentricity then increase rapidly with time (phase II). 
 Finally, in phase III, the eccentricity either  continues to
increase,  but more slowly than in phase II, or
remains constant for $\gamma=1.5$.

\begin{figure*}
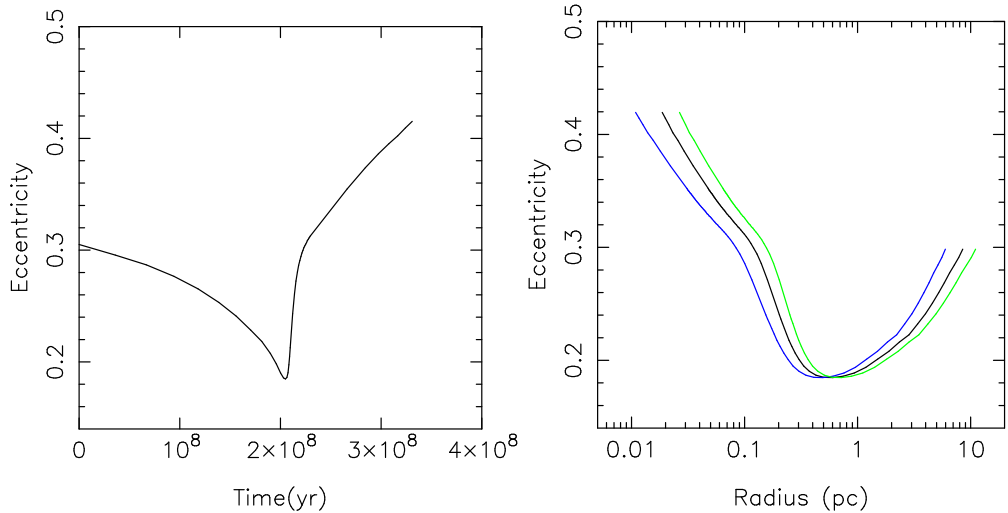

\begin{center}
 \includegraphics[angle=0,width=2.67in]{fig6a.eps}  \includegraphics[angle=0,width=2.51in]{fig6b.eps}
  \caption{ 
Left panel :  eccentricity evolution for a $2\times 10^3 {M_{\odot}}$ black hole in a model with $\gamma=0.8$. 
The initial  apoapsis and periapsis of the orbit are  $12$ and $7$pc respectively
which give a semimajor axis $a\approx 9$pc.
Right panel:  eccentricity versus semimajor-axis (black line),  apoapsis (green line)
  and periapsis (blue line).      
  }
\label{eev}
\end{center}
\end{figure*}

This evolution can  be  understood by considering the changes  of $r_{\rm ap}$
 and $r_{\rm per}$ with time.
 In phase I, the black hole periapsis is close to the core radius  where the difference
 between the density models is small. As a consequence, the eccentricity evolution
is nearly independent of $\gamma$ and the orbits circularize.
 In phase II, $r_{\rm per}$ is well inside the core where the   
smaller dynamical friction results in a rapid eccentricity increase.
Finally, in phase III, the orbit lies entirely inside the core.
As a consequence of the declining dynamical friction at  $r_{\rm ap}$  
the eccentricity growth slows down.
As predicted, for $\gamma=1.5$, the eccentricity remains unchanged in this phase.

These results show that, in the presence of a flat ($\gamma\lesssim 1$) density 
profile, a second black hole found initially on an eccentric orbit can acquire 
very large eccentricities ($\lesssim 1$) before entering the regime where relativistic effects
become important.  In Section \ref{dfde} we discuss in more detail how  very large eccentricities 
may modify the expectations for the GW signal from massive black hole binaries for 
  proposed space-based interferometers.

 In the first phase, when the periapsis is still outside the core,
 the orbit evolves completely in the outer cusp ($\gamma_e=1.8$).
 Evolution in this regime could  lead to a rapid circularization before the black hole reaches the inner core.
 To quantify the amount of circularization 
 in this phase  we computed a further orbit in the model with $\gamma=0.8$,
 adopting  initially  a larger  semi-major axis ($a\sim10$pc) and a smaller eccentricity ($e=0.3$). 
The results of this integration (Figure~\ref{eev}) show that the eccentricity reaches  a minimum value, $e\approx 0.15$,  and then increases rapidly reaching $e\approx 0.3$ at $r_{\rm per}=0.1{\rm pc}$.
At the end of the integration the orbit retains therefore a substantial eccentricity ($\sim 0.4$), even though it  was almost circularized at the beginning of phase II.

  \begin{figure}[b!]
\begin{center} 
  \includegraphics[angle=270,width=2.6in]{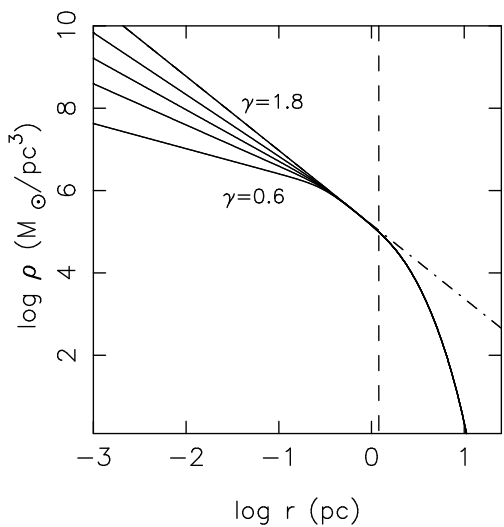}
  \caption{Density profiles of equation (\ref{den2}) with  
  $\gamma=(0.6,1,1.25,1.5,1.8)$,  $r_0=0.3$pc, $\alpha=4$ and 
  truncation radius $r_{\rm  t}=1.2$pc (vertical dashed line). 
The dash-dotted  line gives the coreless model of equation (\ref{fidmod}).
}
\label{mm}
\end{center}
\end{figure}

  \section{$N$-body simulations}
  
The numerical integrations of equation (\ref{dfa}) presented above predict
that a massive body that spirals in to the center of a galaxy containing
a SMBH, and a nuclear star cluster with flat ($\gamma\lesssim 0.6$)
density profile,
 will stall, at a radius that is roughly the core radius. 
Moreover, its  eccentricity is expected to increase steeply once 
the orbital periapsis lies inside the core.
Here we use $N$-body simulations to test these predictions.

 \subsection{Initial Conditions and Numerical Method}
In order to generate equilibrium $N$-body models of
the GC region that extend self-consistently to the 
Sgr A*  influence radius ($r_{\rm bh}\approx 2.5{\rm pc}$)  we  used 
 the truncated mass model
\begin{equation}
\label{den2}
\rho(r)=\rho_0 \left(  \frac{r}{r_0} \right)^{-\gamma}
 \left[
1+\left( \frac{r}{r_0}
\right)^{\alpha}\right]^{(\gamma-\gamma_{e})/{\alpha}}
\zeta(r/r_{\rm t}),
\end{equation}
with  truncation function
\begin{equation}
\label{cut}
\zeta(x)=\frac{2}{\mathrm{sech}(x)+\mathrm{cosh}(x)}.
\end{equation}
With this choice,  
the  density falls off exponentially
 at large radii (i.e., $r>r_{\rm t}$) while
for $r\ll r_{\rm t}$, where 
$\zeta(x) \approx 1-x^4/8$, 
 the model  reproduces almost exactly the density
 of equation (\ref{den}).
As above, we chose  $r_0=0.3$pc, $\alpha=4$, $\gamma_{e}=1.8$ and $\rho_0=1.3 \times 10^6 {M_{\odot}}$.
Monte-Carlo initial positions and velocities were then generated by numerically solving  equation (\ref{fvr}); we stress that the equilibrium models so produced include
self-consistently the effects of the gravitational force from the stars.
Figure~\ref{mm} shows the truncated density profiles for different values
of $\gamma$ and  $r_{\rm  t}=1.2$pc.

The initial conditions were evolved using the direct-summation code $\phi$GRAPE \citep{H07}
which uses a fourth-order Hermite integrator with a predictor-corrector scheme 
and hierarchical time steps.
The performance and accuracy of the code depend both on the time-step parameter $\eta$ and 
on the  smoothing length $\epsilon$. In what follows, we set $\eta=0.01$ and $\epsilon=5\times 10^{-4}$pc.
With these choices, energy conservation was typically of order $0.1\%$ over the entire length of the integration.
Most of the $N$-body integrations were carried out on the 32-node GRAPE  cluster 
at the Rochester Institute of Technology. In addition, a few were carried out in
serial mode using a {\sc Tesla C870} graphics processing unit with {\sc sapporo}, a {\sc cuda} library 
that emulates double-precision force calculations on single precision hardware \citep{sap}.

 \begin{table*}
 \begin{center}
\caption{ \label{t1}} 
\begin{tabular}{lllllllllll}
 \hline
 ${\rm Model}$  & $\gamma$  & $N$ &  $r_{\rm t}$ &  $M $  &  $m$ & $e_{\rm in}$  &   $r_{\rm in}$ &  $r^*$ &ln$\Lambda$ \\ 
 & \phantom   &k  & (pc)& $(10^3{M_{\odot}})$ &  (${M_{\odot}}$) &  &  (pc)&(pc) \\
 \hline 
 A1      & $0.6$ &  230 & 1.2     &  5			&  22		&  0		& 1 & 0.07&6.7\\
 A2      & $0.6$ &  130  & 1.2    &  5 	 		&  38		&  0		& 1  & 0.07 &6.6\\
 B1      & $0.8$ &   230 & 1.2    &  5			&  22 	&  0	& 1        &  0.06 &    6.9  \\	
 B2      & $0.8$ &   130 & 1.2    &  5			&  38 	&  0	& 1      & 0.06    &    6.9 \\	
C        & $0.6$ &   80   & 0.6    &  5				&  26		&  0	& 0.5 & 0.07  &   6.3\\
 \hline
 D        & $0.6$ &  130 & 1.2   &  2		& 38		&  0		  & 0.3	& 0.05 & ...  \\
 E       & $0.6$ &  130 & 1.2   &  10 	& 38		&  0	            	  & 1	& 0.10 &  6.4 \\
 F        & $0.6$ &  130 & 1.2   &  50	& 38		& 0	  	         	 & 1	&    0.18       &4.8\\
 \hline
 G1       & $0.6$ &  200 & 1.2   &  5	& 25		&  0.54	  & 1 & 0.07 &6.9\\
 G2       & $0.6$ &   100 & 1.2  &  5	& 50		&  0.54	  & 1 &  0.07	 &6.9\\
 \hline
 \\ \\
\end{tabular}
\end{center}
\end{table*} 

Table 1 gives the parameters of the $N$-body models.
The initial distance of the secondary black hole is given by $r_{\rm in}$ while
its initial orbital eccentricity  is $e_{\rm in}$.
The quantity $r^*$  is the radius at which the initial mass in stars equals   $M$,
the mass of the second black hole.
All of our $N$-body models had $r_{\rm in}<r_{t}$, so that the 
orbital evolution is expected to be very similar to that in the corresponding non-truncated models.
In order to study the dependence of the results  on the secondary black hole mass
we run simulations with a range of masses,
 $M=(2000, 5000, 10000, 50000) {M_\odot }$.
Two  cases with nonzero initial eccentricities (runs G1 and G2, with
$e_1=0.54$) were also considered.

 \subsection{The Coulomb Logarithm}

In  Table 1 we report the values of the Coulomb logarithm
extracted from each $N$-body integration.
The value of  ln$\Lambda$ was obtained by minimizing the quantity: 
\begin{equation}
\sum^n_{i=1} \left[r_i(t)- r'(t,{\rm ln} \Lambda) \right]^2,
\end{equation}
outside  a galactocentric radius $r>0.3$pc. 
Here,  $n$ is the number of $N$-body data points, 
$r_i(t)$ is the position of the black hole in the $N$-body simulation
at time $t$, and  $r'(t)$ is its position at the same time 
evaluated by means of the  Chandrasekhar's formula (\ref{dfa}).
Since analytical expressions are not available for the 
trajectory of an inspiraling black hole,  in order to obtain the expected position  $r'(t)$ 
at any given time, we first solved numerically the equations of motion (\ref{em})
 and then built a spline interpolant from the results of the integration.
  This procedure was applied only in the part of the orbit outside the core, where 
 equation (\ref{dfa}) is able to  describe accurately the black hole orbit. 
In this way, unlike in most previous studies, we could obtain an estimate of the Coulomb logarithm 
without making any assumptions about the velocity distribution of the field stars (e.g., that it followed a Maxwellian  distribution).

Our simulations do not show any obvious dependence of ln$\Lambda$ on either 
the number of particles  or on the initial eccentricity. 
We found an average value of ln$\Lambda = 6.5 \pm 0.2$,
in essentially perfect   agreement with the value 
reported by \citet{Spinnato}: ln$\Lambda = 6.6 \pm 0.6$.

\begin{figure*}
\begin{center}
  \includegraphics[angle=270,width=5.9in]{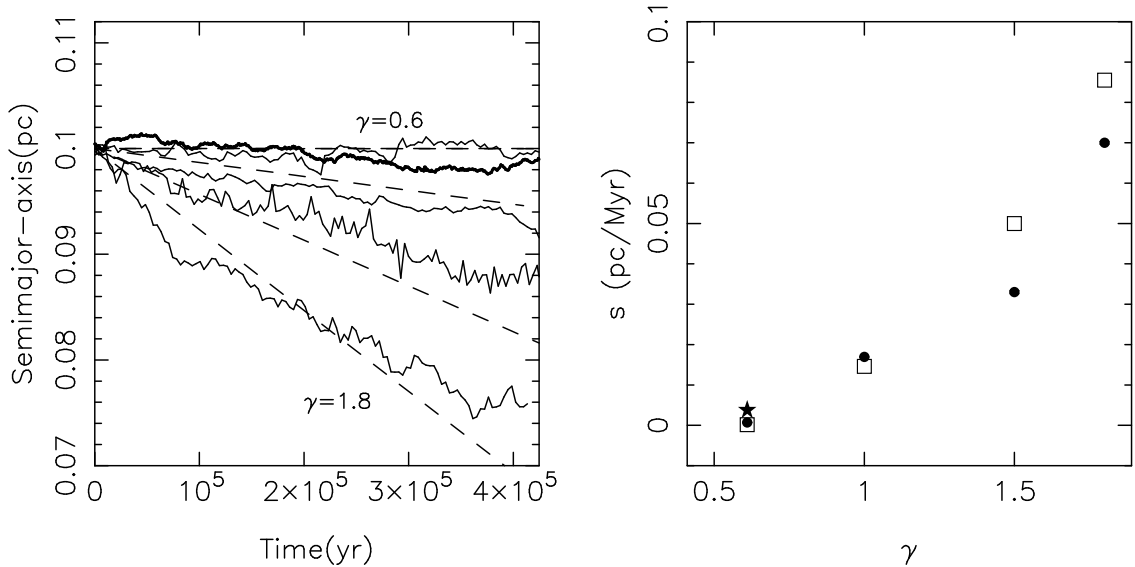} \\
  \caption{  Left panel: evolution of the semi-major axis for a $5000 {M_\odot }$ black
    hole in the short $N$-body integrations,
  for different values of the central density slope (from top to bottom, $\gamma=0.6,~1,~1.5,~1.8$).  
  The thicker line is from the high-$N$ integration,  with $N=500,000$ and $\gamma=0.6$.
  Dashed lines are predictions from Chandrasekhar's formula (\ref{dfa}) using ln$\Lambda=6.6$. 
  For $\gamma=0.6$ there is no significant evolution of the orbit in the considered 
 interval of time.
  Right panel:  orbital inspiral rates $s=-da/dt$ computed for the simulations displayed on the left panel 
    as a function of  $\gamma$ (filled circles).  Open squares give the predictions from
  Chandrasekhar's formula.  The star symbol is the decay rate computed from the 
  high resolution run ($N=500,000$ and $\gamma=0.6$).
    }
\label{D0}
\end{center}
\end{figure*}

\subsection{Results}
\subsubsection{Circular Orbits} \label{NBR}

The first simulations  we performed consisted in evolving the massive body on a circular orbit with initial radius 
$0.1$pc  (i.e., smaller than the stalling radius when $\gamma\lesssim 0.6$) and for a  time corresponding approximately  to 
300 orbits (i.e., $\sim 4\times 10^5$yr at this distance). We used $N=130,000$,   $M=5000{M_{\odot}}$  and $\gamma=(0.6,~1,~1.5,~1.8)$.
We also implemented a high-resolution simulation with $N=500,000$ for the model with  $\gamma=0.6$.
As in most of the longer simulations of Table 1, the truncation radius was   $r_{\rm  t}=1.2$pc.
These shorter integrations allowed us to study  dynamical  friction, while limiting the deviations of the models from 
their initial configuration that was found to occur on longer timescales as a result of two-body relaxation and perturbations
from the massive object (see below).
The eccentricity of the orbit remained small during these integrations ($e\lesssim 0.1$).

\begin{figure}
\begin{center}
  \includegraphics[angle=270,width=2.3in]{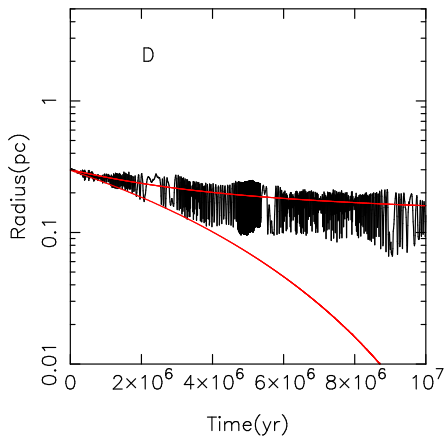} \\
  \caption{  Trajectory of a $2000 {M_\odot }$ black hole into  a core  with  $\gamma=0.6$ (model $D$). 
  The top-red  line is  the theoretical prediction obtained from Chandrasekhar's formula (\ref{dfa}) using ln$\Lambda=6.6$. 
 The bottom red curve shows the predicted inspiral in a $\gamma=1.8$ cusp.}
\label{D1}
\end{center}
\end{figure}

Figure~\ref{D0}  shows the time evolution of the  semi-major axis of the orbits and the 
rate of orbital decay $s=-da/dt$ as a function of $\gamma$.
The agreement with the decay rate computed using Chandrasekhar's formula  (\ref{dfa})  (open squares) is good.
For $\gamma=0.6$, there is not  any significant evolution of the orbit in the considered interval of time
 and, consequently,  $s\approx 0$.

A similar conclusion is implied by  Figure~\ref{D1} which shows the trajectory  of a $2000{M_{\odot}}$ black hole in model D, a longer integration with
$N=130000$ and $\gamma=0.6$.
 Initially, the black hole sinks rapidly to the center,  reaching  $\sim r_{\rm st}$ in $\sim 3$Myr.
 As the inspiral progresses, the orbit becomes more eccentric ($e \approx 0.3$ at $4$Myr).
  At later times ($\gtrsim 4$Myr), the orbit shows no sign of further decay, oscillating  in radius between 
  $\sim 0.1$ and $\sim 0.2$pc.
The orbital eccentricity remains almost constant in this phase.

 These findings, obtained for a flattened density cusp around a SMBH,  seem to confirm the theoretical predictions made above:
i) dynamical friction ``vanishes" within $r_{\rm st}\approx0.15$pc; ii) the orbital eccentricity of an infalling body 
increases with time.

 However,  in any $N$-body simulation,  stars are
continuously scattered by gravitational encounters with other stars, 
with the result that the initially empty phase space region responsible
for the vanishing dynamical friction force will gradually be filled.
 In addition, due in part to the low central density of our GC models when
$\gamma$ is small,
the radius at which  the cumulative mass in stars becomes comparable to that of the inspiraling black hole 
can be of order $r_{\rm st}$,  even for relatively small  $M$ (see table 1). 
$N$-body simulations have shown that, in these circumstances,    
the  orbit deviates  from the theoretical prediction 
of the Chandrasekhar's formula as a consequence of perturbations induced
by the infalling black hole on the inner cusp \citep{BGP:06,LB:08}.
 Finally, it is not clear whether the approximations made in deriving equation (\ref{dfa}),
which was the basis for the red  lines plotted in Figure~\ref{D1}, are 
reasonable, or how large might be the frictional force from fast moving stars that populate the low density core.
In fact, as we now demonstrate, these additional effects have a substantial influence
on the long-term evolution of the black hole orbit.

\begin{figure*}{ ~}
\begin{center} 
	$\begin{array}{ll}  
  \includegraphics[angle=270,width=2.44in]{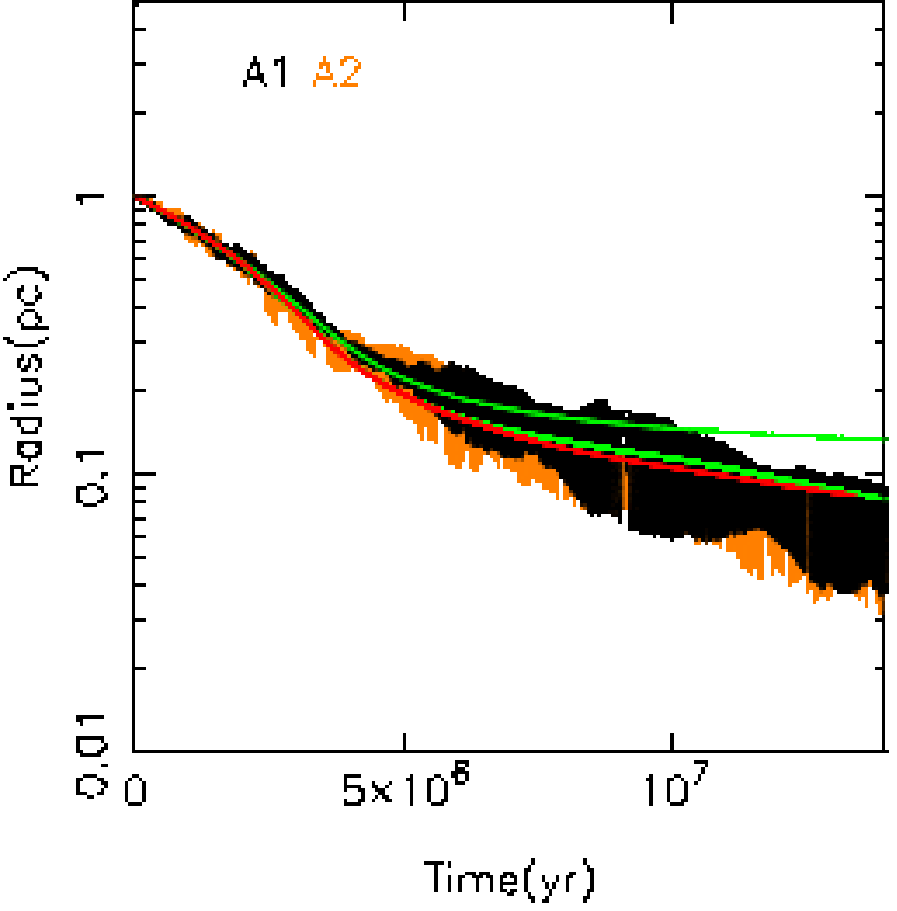} &  \includegraphics[angle=270,width=2.348in]{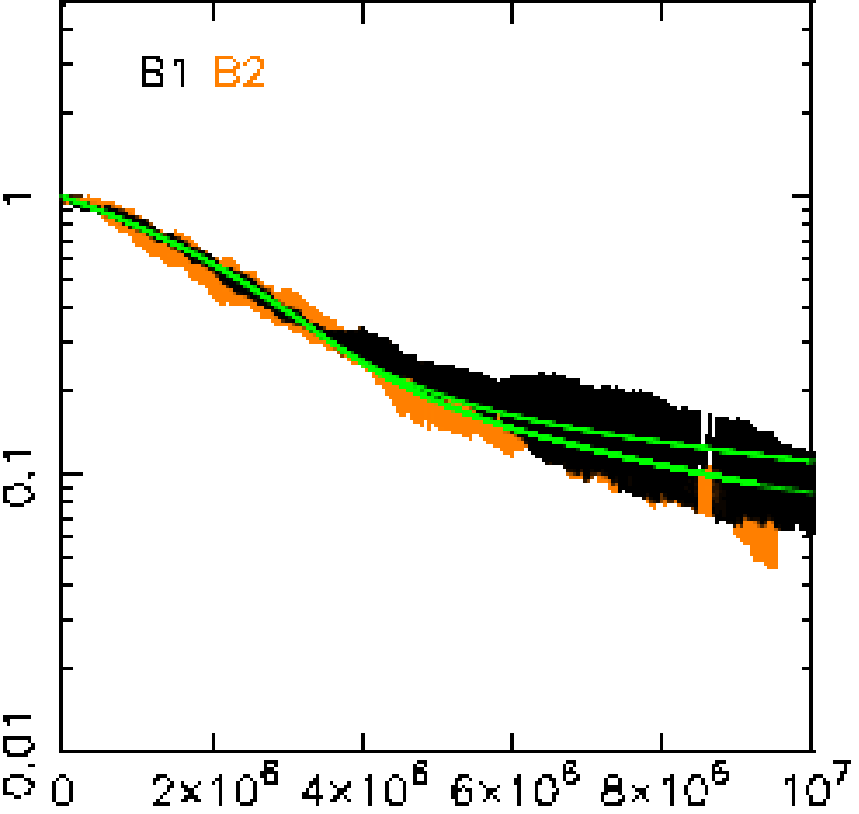} \\ 
  \includegraphics[angle=270,width=2.42in]{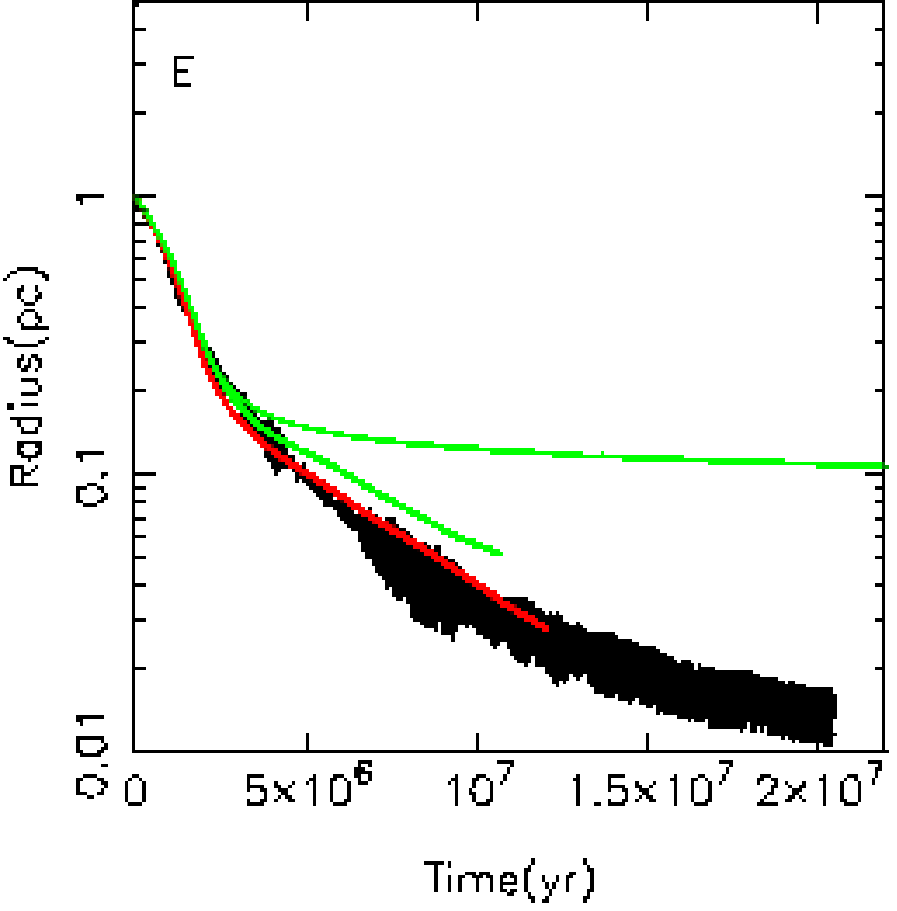}  &  \includegraphics[angle=270,width=2.255in]{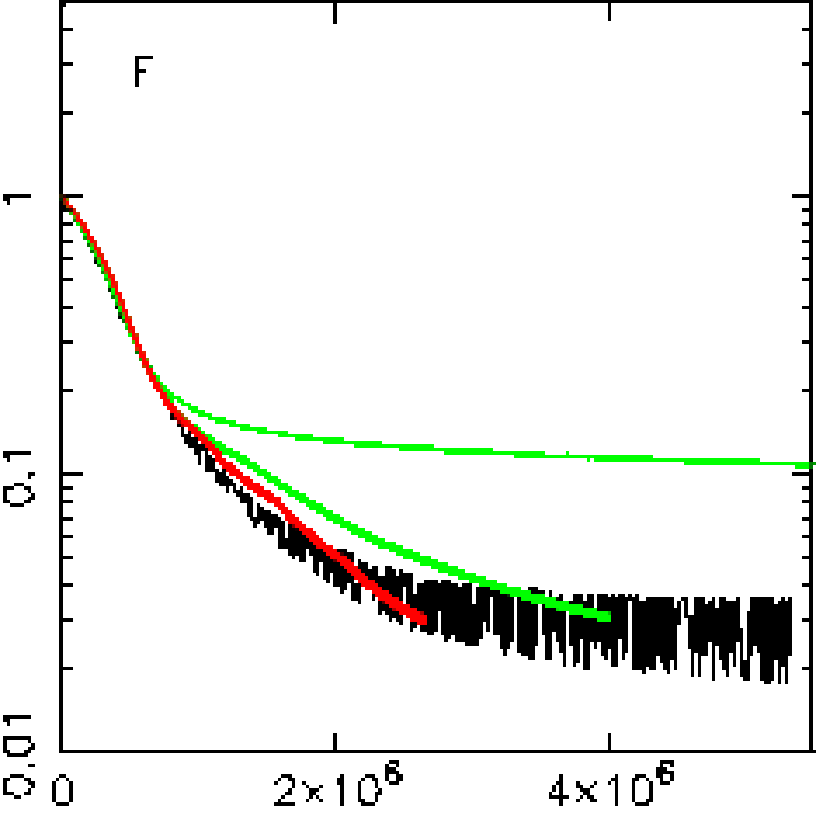} \\
    \end{array}$
  \caption{   Orbital evolution  of the  second black hole in models  A1, A2, B1, B2, E and F.
   Solid green lines show predictions  assuming a fixed background of stars.
   Upper green curves are obtained by using the standard  Chandrasekhar's formula  (i.e., equation~(\ref{dfa})),
   while lower green curves    give the orbital decay computed using equation (\ref{dfatot}) with $p_{max} =0.5~{\rm pc}$.
    Red lines   were obtained   with   equation (\ref{dfatot})  but allowing $f(v_\star)$ and $\rho(r)$ to change according to the evolution   of the $N$-body system. 
  	}
\label{r1}
\end{center}
\end{figure*}

Figure~ \ref{r1} shows the trajectory of the black hole for some of the $N$-body 
integrations from Table 1  and compares them to the evolution predicted by 
Chandrasekhar's formula   (\ref{dfa}) (upper green  curves). 
(In the upper panels, the comparison is displayed only for the higher resolution runs, i.e., models  $A1$ and $B1$.)
Although  the agreement with the theoretical prediction appears fairly  good, at least for $M= 5000 {M_{\odot}}$,
when $\gamma=0.6$, the $N$-body integrations reveal a faster decay than predicted.
Either some the frictional force must come from stars with velocities $v_\star>v$, or  the background stellar 
distribution is changing during the  inspiral (or both). These two possibilities are investigated in what follows.

\emph{Dynamical friction from fast-moving stars.}
Equation (\ref{dfa}) was derived under standard approximations that ignore the contribution 
from non-dominant terms and the velocity dependence of ln$\Lambda$.
Although  these approximations are reasonable when there is a large  fraction of 
stars with low velocities (i.e., $v_\star<v$), it is unclear whether they can be applied to a region
populated mostly by stars  moving faster than the black hole.

Without these assumptions, the  instantaneous
dynamical friction acceleration becomes \citep{CH:43}:
\begin{eqnarray}\label{dfatot}
\boldsymbol{f_{\rm fr}} &=&-4\pi G^2  M \rho(r) \frac{\boldsymbol{v}}{v^3} \int^{\sqrt{-2 \phi(r)}}_0 dv_\star 4\pi f(v_\star) v_\star^2  \nonumber \\
&&\times \frac{1}{8v_\star}  \int^{v+v_\star}_{|v-v_\star|} dV \left( 1+ \frac{v^2-v_\star^2}{V^2}  \right) {\rm ln}\left( 1+\frac{p_{\rm max}^2 V^4}{G^2 M^2} \right)~,~~~~
\end{eqnarray}
where $f(v_\star)$ is the velocity distribution of field stars, and $p_{\rm max}$ is the effective, maximum value of the impact parameter.
In this more accurate treatment, some of the dynamical friction force 
is due to stars moving more rapidly  than the massive particle \citep{CH:43,W:49,M:01}.
If the condition $p_{\rm max} V^2/G M\gg1$ is satisfied, 
the frictional force can be approximated as (Chandrasekhar~1943, equation 30):
\begin{eqnarray} \label{dfa3}
\boldsymbol{f_{\rm fr} }  &\approx& \boldsymbol{f^{(v_\star<v)}_{\rm fr}+f^{(v_\star>v)}_{\rm fr}}     = -4\pi G^2  M \rho(r) \frac{\boldsymbol{v}}{v^3}\nonumber \\
&&\times {\Big(} \int_{0}^v dv_\star 4\pi f(v_\star) v_\star^2  ~{\rm ln}  
\left[ \frac{p_{\rm max}}{GM} \left( v^2-v_\star^2 \right)	 \right]   \\
&& +\int^{\sqrt{-2 \phi(r)}}_v dv_\star 4\pi f(v_\star) v_\star^2    \left[ {\rm ln} \left( \frac{v_\star+v}{v_\star-v} \right)-2\frac{v}{v_\star} \right]  \Big). \nonumber
\end{eqnarray} 
Inside $r_{\rm st}$, dynamical friction is produced mostly  by stars with $v_\star>v$ and  the first term in the integral becomes negligible.
This  shows the weak dependence of the frictional deceleration inside the core  on $p_{\rm max}$.

\begin{figure*}{ }
\begin{center}  
\includegraphics[angle=270,width=5.5in]{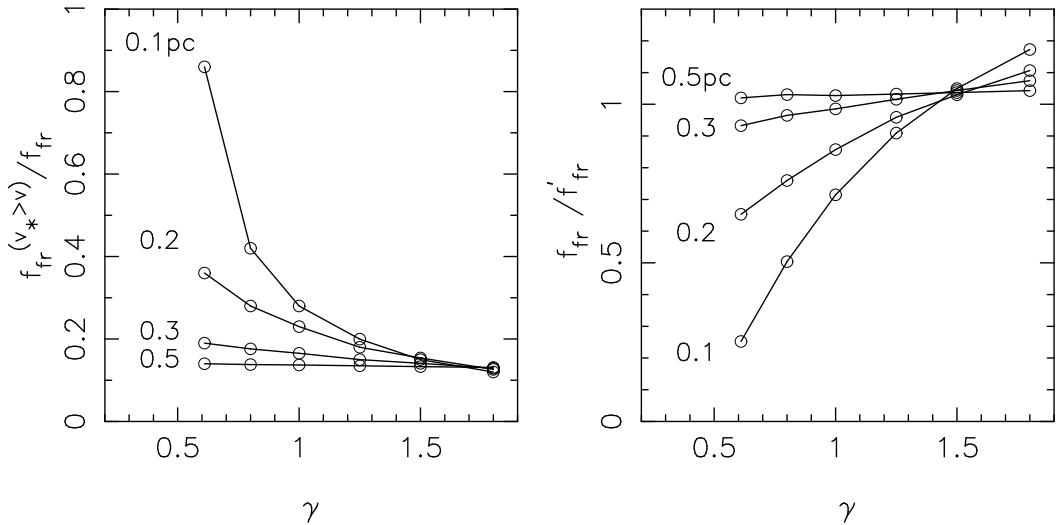}
  \caption{ Left panel: fraction of the dynamical friction force that is predicted to come from stars with $v_\star>v$ as a function  of $\gamma$, 
  at different galactocentric radii: $r= 0.1, 0.2, 0.3$  and 0.6 pc.	 Equation  (\ref{dfatot}) was used to compute these curves.
 When  $\gamma=0.6$, dynamical friction at small radii comes only from stars with  $v_\star>v$. 
 As either $\gamma$ or $r$ increase, the contribution from fast moving stars decreases.
   Right panel: total dynamical friction force in units of the frictional deceleration computed assuming a Maxwellian distribution of velocities.
  The frictional force produced by stars with $v_\star>v$, in the flattened cusp (i.e., $\gamma=0.6$ and $r \lesssim 0.2$pc)  is much smaller than that
  obtained under the simple assumption of thermal distribution of velocities.
  In both panels we adopted $p_{\rm max}=0.5$pc and $M=1000{M_\odot }$. In the right
  panel, we used ln$\Lambda=6.6$ to solve equation (\ref{ff}).  
   }\label{nct2}
\end{center}
\end{figure*}

Adopting equation  (\ref{dfatot}), 
with $p_{\rm max}=0.5$pc, for the frictional
force that appears in the equations of motion (\ref{em}), 
we obtained the lower green curves in Figure~\ref{r1}, 
which show much better agreement with the $N$-body results.
Evidently, the standard expression for dynamical friction,
equation~(\ref{dfa}) , is inadequate to describe the orbital evolution of a massive body at the GC
in the case that the density profile of the nuclear star cluster is shallow.
This is apparently a consequence of neglecting the non-dominant terms, and not, for instance, of the  
 assumed independence  of the Coulomb logarithm on the field star velocity distribution. 
For models A1 and A2, Lagrangian radii showed essentially no evolution, indicating
the absence of any significant change in the  stellar distribution induced by the
second black hole. 
We conclude that (at least)  some of the drag within $r_{\rm st}$  is due to field stars  with $v_\star>v$.
 The red lines in Figure~\ref{r1} were derived from
equation   (\ref{dfatot}) but using  a time dependent distribution  function $f(v_\star,t)$ extracted  (at time $t$) from the $N$-body models (see below).
For  models A1 and A2 the red curves agree  exceptionally well with the $N$-body results and they essentially 
match the results of the semi-analytical integration that takes into account the friction from fast moving stars.
We conclude that for these runs  it would be appropriate to ignore the influence of the second black hole 
on the stellar distribution.

In  the left panel of Figure~\ref{nct2}  we plot the fraction of the dynamical friction force 
that is predicted, by equation~(\ref{dfatot}), to come from stars with $v_\star>v$, for different values of the inner cusp slope and at different radii. 
In the right panel of the figure, the total frictional deceleration in our models is given  in units of the frictional force
computed under the assumption of a Maxwellian  distribution of velocities:
 \begin{eqnarray}
\boldsymbol{f'_{\rm fr}}=\frac{-4\pi G^2  M \rho(r) {\rm ln} \Lambda}{v^3}\boldsymbol{v}  \left[{\rm erf}(X)-\frac{2X}{\sqrt\pi} e^{-X^2}\right],~~~~ \label{ff}
\end{eqnarray}
with $X=v/\sqrt2\sigma$.
Clearly, this equation, often used in the past to describe the orbital evolution of a massive object into the GC, 
overestimates the frictional drag  within $r\lesssim 0.2$pc for $\gamma \lesssim 1$.

\begin{figure*}
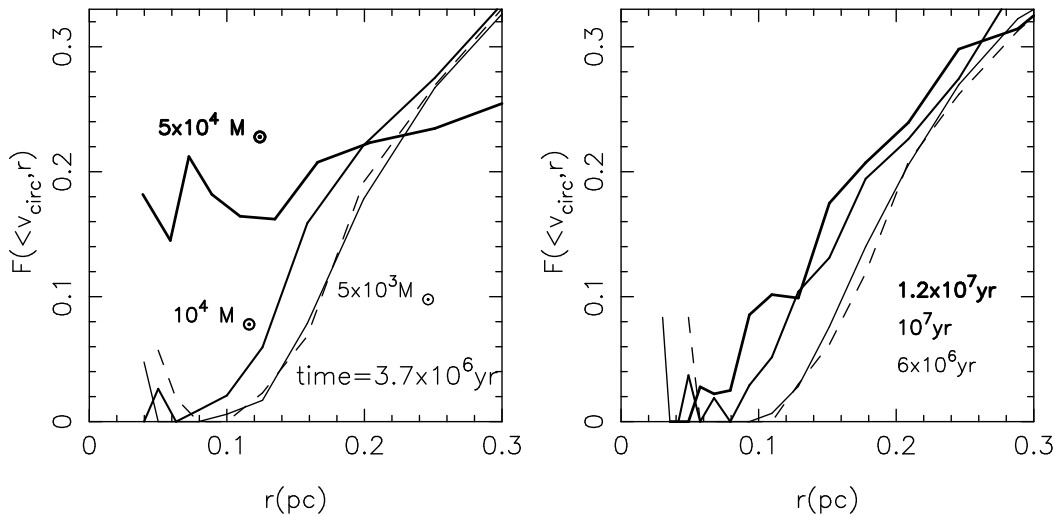

\begin{center}
$\begin{array}{cc}  
  \includegraphics[angle=270,width=2.7in]{fig12a.eps} &
   \includegraphics[angle=270,width=2.7in]{fig12b.eps} \\
                     \end{array}$           
  \caption{ Left panel:  Fraction of stars with velocities less than the local circular velocity $F(<v_{\rm circ},r)$
   as a function of radius, at the same time
($3\times 10^6$yr) for models A2 ($M= 5000 {M_{\odot}}$), E ($10000{M_{\odot}}$) and F ($50000{M_{\odot}}$) .  The dashed curve corresponds to the initial 
configuration.
The larger the mass  of the black hole the faster the
changes of the model in  velocity space.
Right panel: $F(<v_{\rm circ},r)$ as a function of radius for model A1 at different times.
Due to two-body relaxation, stars are scattered toward low velocities and the hole in phase space that 
characterized the initial configuration is gradually filled up.  }
\label{VS}
\end{center}
\end{figure*}

\bigskip
\emph{Influence of the second black hole on the field-star distribution.}
For larger masses of the infalling body, i.e. $M \gtrsim 10000{M_{\odot}}$, the  perturbations which it induces in  the background system
introduce a complex time dependence of the phase-space distribution.
During the orbital inspiral,  the black hole scatters stars into the inner cusp;
consequently, once it reaches  $\sim r_{\rm st}$, it will ``see" stars with $v_\star<v$
that contribute to the frictional acceleration from that point on.  

In order to  test Chandrasekhar's  formulae under these circumstances, 
the  black hole equations of motion were integrated in a time-varying potential whose properties were 
varied over time in a way designed to mimic the evolving $N$-body models.
In more detail, the density of the $N$-body model was computed at fixed intervals of time 
 by binning particles in concentric logarithmically-spaced shells.
At the same time the  velocity distribution of  field stars was obtained directly from 
the $N$-body model.  Finally, the  black hole equations of motion were numerically integrated 
as described in section (\ref{co}) using expression (\ref{dfatot}).
In this way,  we were able to approximately account for the back-reaction of the 
second black hole on the stellar distribution.
It is worth noting that, even with this more sophisticated approach, two relevant assumptions are retained:
i) any induced deviation of the models from isotropy is neglected;
ii)  the black hole is assumed to move always on a circular orbit, while the
$N$-body simulations clearly show  an increase of the orbital eccentricity with time.
The  red curves   of Figure~ \ref{r1}, obtained through this numerical procedure, show that
even when the galactic nucleus is  rapidly deviating from its initial configuration,
  Chandrasekhar's theory  can still  accurately reproduce  the $N$-body results if
the changes in the stellar distribution   are taken into account and the fast moving stars are included
when computing the frictional force.

  \begin{figure}
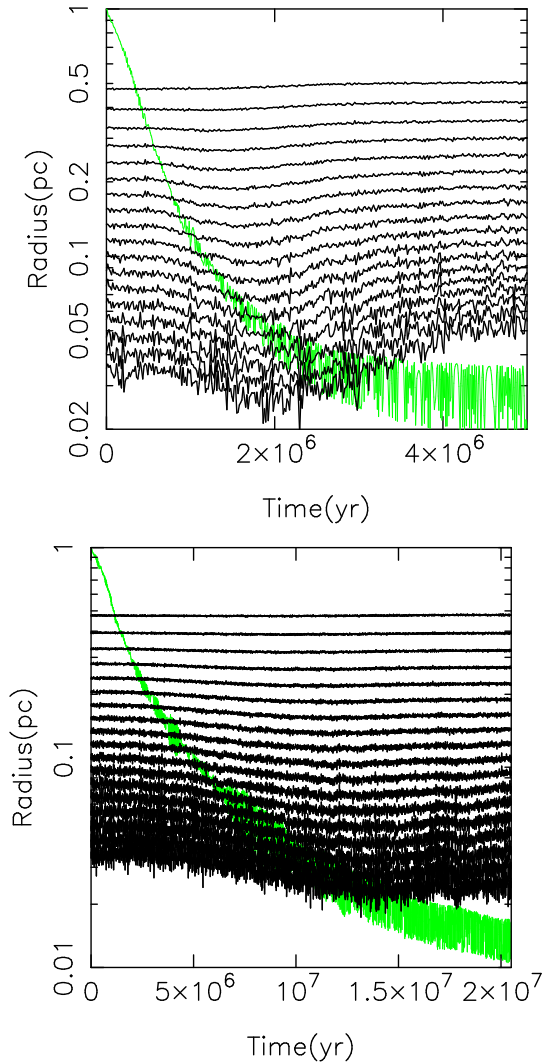

\begin{center}
$\begin{array}{cc}  
  \includegraphics[angle=270,width=2.65in]{fig13a.eps} \\
     \includegraphics[angle=270,width=2.81in]{fig13b.eps} 
                     \end{array}$           
  \caption{ Lagrangian radii evolution of models F (upper panel) and E (lower panel).
  Green curves show the  position of the massive body.  }
\label{CS}
\end{center}
\end{figure}

In Figure~ \ref{VS} we show the evolution induced by the second black hole in the
velocity distribution of the model, by plotting the function
$F(<v_{\rm circ},r)$ at the same time ($3 \times 10^6$yr)  for different masses (left panel). 
In addition, we show how $F(<v_{\rm circ},r)$, for  $M= 5000{M_{\odot}}$, evolves
as a function of  time (right panel).  In this latter case
two-body relaxation causes  the diffusion of stars at low velocities and
the stalling radius  is shifted from the initial $\approx0.1$pc to $\approx0.05$pc by the end of the simulation.
 We note that -- in a real galaxy with much larger $N$ -- this effect
would be essentially absent.

\begin{figure*}
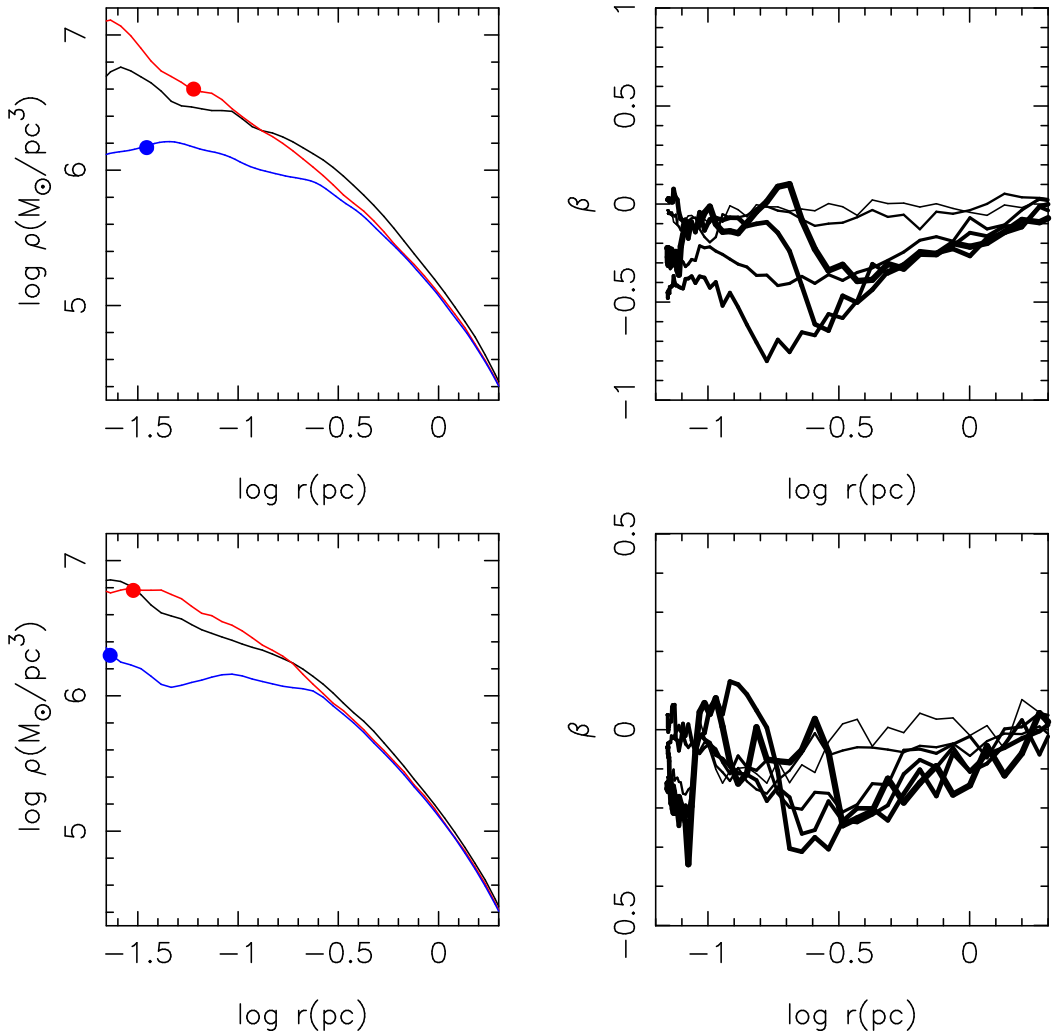

\begin{center}
 \includegraphics[angle=0,width=5.5in]{fig14a.eps} 
 \includegraphics[angle=0,width=5.5in]{fig14b.eps} 
   \caption{ Left panels: density profile evolution  in run F (upper panel) and E (lower panel). The black curve corresponds to the initial model;
   the red line  is obtained at time $10^7$yr for run E and at   $2\times 10^6$yr for run F, while the blue lines are  the density profile of the final models, after
   the secondary black hole has stalled carving  out a deficiency of stars in the inner regions. Filled circles indicate the position of the 
   inspiraling .   Right panels: Evolution of the anisotropy parameter in the models. Line thickness increases with time. As the black hole spirals in, 
   it induces tangential anisotropy in the background system. }     
  \label{DLM}
\end{center}
\end{figure*}

 Figure~ \ref{CS} illustrates the changes in the configuration-space density for models E and F via the time evolution of their Lagrange radii. 
The time evolution of models E and F is remarkable:
in model F, the   perturbations on the stellar distribution are initially so large that the core fills up during the first $\sim 2 \times 10^{6}$yr.
At this point,  the black hole, at a galactocentric distance of $\sim 0.05$pc,  starts to carve out  the inner region, destroying  the cusp that it  created before.
The final model has a core of size $\sim 0.2$pc and the internal slope is $\gamma \lesssim 0.5$. 
However its density is, everywhere within $1$pc, 
smaller than that of the initial model as a consequence of displacement of stars from the cusp.
A qualitatively similar evolution was found in model E.
Figure~\ref{DLM}  shows the induced evolution of the density profile for runs E and F as well as the time variation of the anisotropy parameter,
defined as
\begin{equation}
\beta=1-\sigma_t^2/\sigma_r^2~, \label{eq:beta}
\end{equation}
with $\sigma_t$ and $\sigma_r$ tangential and radial velocity dispersions respectively. 

In summary,  a straightforward interpretation of our $N$-body results is that equation  (\ref{dfa})  reproduces remarkably well
the real decay rate of a massive object into the GC  only until it reaches the stalling radius. In the subsequent evolution,
the orbital decay slows down as a consequence of the lack of slow moving stars in the inner galactic nucleus (see Figures~\ref{D0} and~\ref{D1}), but 
it never drops  to zero, due apparently to the frictional  force generated by stars moving faster than the  inspiraling black hole ( Figures~\ref{r1} and~\ref{nct2}).

A massive body  of mass  $M \approx 1000{M_{\odot}}$, starting from distances of order $r_{\rm bh}$, will reach 
a galactocentric  radius  $\sim 0.01$pc in $\sim 10^8$yr. 
 For larger masses (i.e., $M \gtrsim 10000{M_{\odot}}$), during the inspiral, the black hole
 enhances the diffusion of stars into the phase-space region that was initially nearly
empty (Figures~\ref{VS} and~\ref{CS}).  
During the stalling phase a low density core is rapidly regenerated  by the  second black hole as it displaced  stars from the cusp.
Notice that, in our models the stalling distance is about ten times larger than that found in previous works that assumed a 
collisionally-relaxed, steeply-rising density profile around the central black hole \citep[e.g.,][]{BGP:06,LB:08}.

We note in passing that the background stars have orbital periods similar
to that of the massive body.  It is conceivable that correlations may be induced
by the massive body in the orbital elements of the stars that will change the
evolution significantly away from that produced by an uncorrelated background.
On the other hand, two-body relaxation in the $N$-body models will tend to
de-correlate the background response, leading, perhaps, to a better correspondence
with the predictions of Chandrasekhar's theory.

 \subsubsection{Eccentric Orbits}
 In this section, we investigate the rate of change of the orbital eccentricity as a consequence 
 of dynamical friction.
 We devised two simulations that  differ only in the number of particles: 200,000 and 100,000.
 We refer to these simulations as runs G1 and G2 respectively (see Table 1);
both have $\gamma=0.6$. 
 The black hole was initially placed at   a radius of $r_{\rm in}=1$pc on an eccentric orbit with $e_{\rm in}=0.54$.
 As discussed earlier (Section \ref{ecc}),  when the orbital periapsis lies within  the core,
the orbit is expected   to become more eccentric
as a consequence of the declining   frictional force in this region. 
 
  \begin{figure*}
\begin{center}
$\begin{array}{cc}  
  \includegraphics[angle=0,width=5.3in]{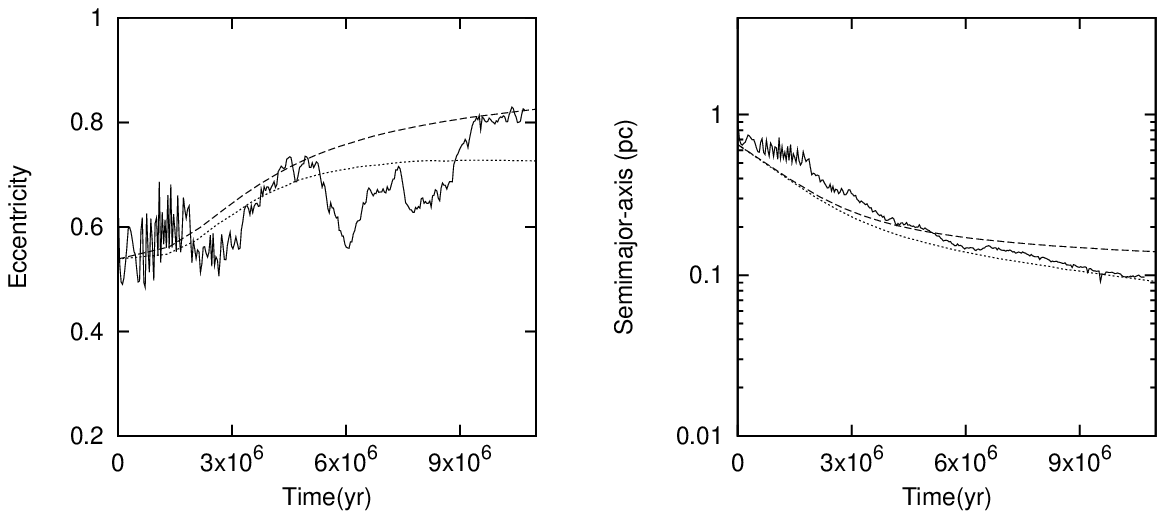} \\
    \includegraphics[angle=0,width=5.3in]{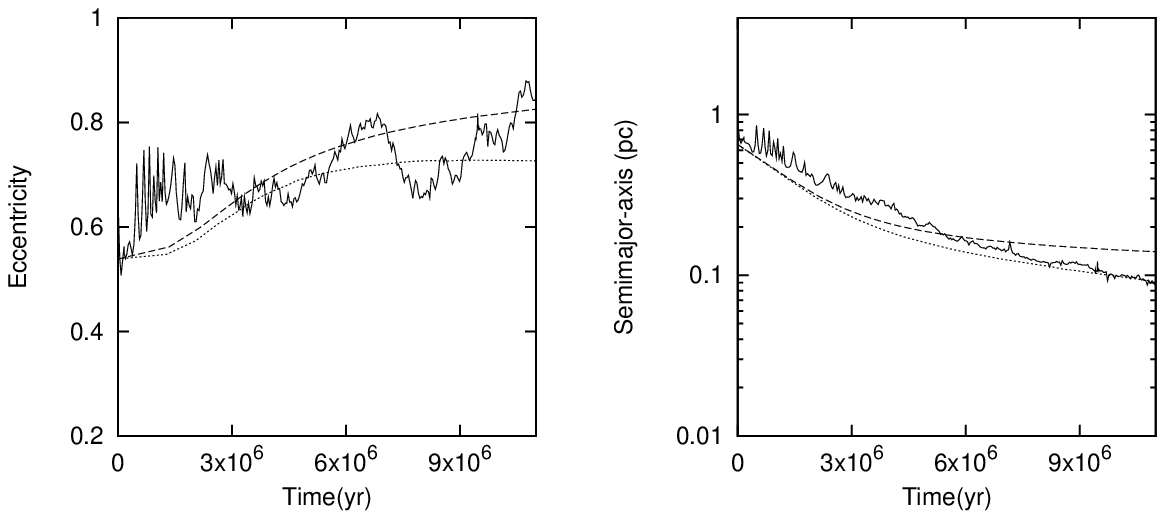} \\
                     \end{array}$           
  \caption{  Evolution of eccentricity and semi-major axis  for models G1 (upper panel)
  and G2 (lower panel) that differ only in the number of field particles: $N$=200,000 and 100,000 for models G1 and G2 respectively.
  Dashed lines are  the  theoretical predictions from equation   (\ref{dfa}). Dotted lines were obtained with  equation  (\ref{dfatot})  (i.e., including
  the  frictional drag from stars with $v_\star>v$), where we
  used $p_{\rm max}=0.5$pc.
   As the black hole spirals in, its orbital eccentricity increases.  This conclusion is quite robust, showing essentially no dependence on
    the number of background particles.
  }
\label{EO}
\end{center} 
\end{figure*}

 Figure~\ref{EO}, shows the evolution of the  eccentricity and semi-major axis of the orbit
 as a function of time, demonstrating that, at least qualitatively,
 Chandrasekhar's theory reproduces the evolution.
Although the eccentricity undergoes significant fluctuations,  it evidently drifts toward 
larger values  with time. 
This behavior  is quite robust showing  a negligible  $N$-dependence.

It is generally assumed that dynamical friction, in power-law density models with an isotropic velocity distribution,   
 would circularize the orbit of an infalling body  (see for instance \citet{BGP:06}).
  Our $N$-body simulations demonstrate that in models characterized by a
  flat density profile and  
  a central SMBH, the eccentricity can instead be an increasing function of time.
  
\section{ Gravitational Wake }
An alternate way to look at dynamical friction is in terms of the acceleration
produced by the overdensity of stars that accumulate behind the massive body --
the ``gravitational wake'' \citep{DC:57,maro,MU:83}.
The expression for the response wake in a homogeneous medium is given for arbitrary spherical density distribution in \citet{w:86}. 
The existence of a wake has rarely been confirmed in $N$-body simulations; an isolated example is provided by 
\citet{wk02} (see also \citet{wk07}) who show  the wake induced in a dark-matter halo 
by a stellar bar.
Other examples include \citet{w:89}, \citet{hw:89}, \citet{vw:00}.
 
\begin{figure*}{ }
\begin{center}  
\includegraphics[angle=270,width=6.5in]{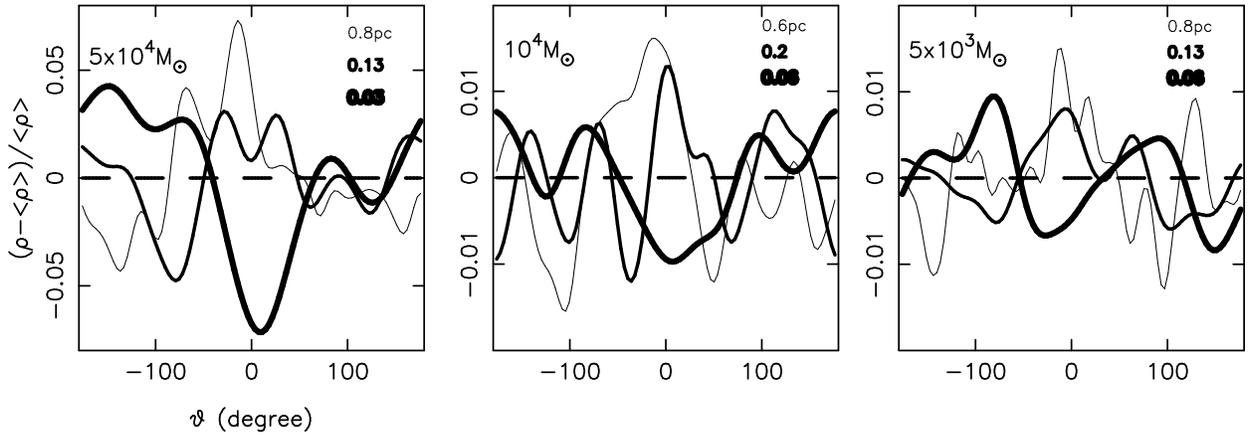}
  \caption{ Relative overdensity  in the $N$-body models for runs A1, E and F along the black hole orbit.
Line thickness decreases with increasing galactocentric distance. 
In the plots, the second black hole is always located at $\theta=0$ with $\dot{\theta}>0$.
}\label{wake}
  \end{center}
\end{figure*}

We searched for the wake in our $N$-body simulations by computing the relative overdensity
at each radius along the orbit of the second black hole.
The $N$-body models were first rotated in such a way that the second black hole was
situated at $y=z=0$ with $v_z=0$ and $v_y>0 $. 
The density at any position was then estimated using a Gaussian kernel with radially-varying smoothing length. 
Figure~\ref{wake} shows the results in runs A1, E and F as a function of  the azimuthal angle  $\theta$ at different radii and for different values of $M$.
In the figure, the black hole lies at $\theta=0$ with $\dot{\theta}>0$ and the average density is defined as $(1/2\pi) \int_{-\pi}^{\pi} d\theta~ \rho(\theta) $ :
outside the core ($r\gtrsim 0.3~$pc),
the peak in the overdensity lies at $-20<\theta \lesssim 0^{\circ}$, 
independent of $M$,
and the amplitude of the overdensity increases with black hole mass, 
as expected. The wake is therefore always just behind the massive body in this phase.
When  $r\lesssim0.3~$pc, for $M=5000-10000 {M_{\odot}}$, the density enhancement is reduced but its position remains essentially unchanged. 
The reduced amplitude  of the wake  inside the core explains why  the frictional force is greatly suppressed in these regions.
For larger masses, the angular dependence of the overdensity in this phase is more complex, revealing, in some cases, two distinct peaks. 
During this phase, the mass distribution is  affected by gravitational scattering
from the massive body.
Finally, when  the black hole is well inside the core,  the density maximum  is  seen to
lie at large angular separations ($\theta \lesssim - 100^{\circ}$) from the black hole. 
Indeed, a  density ``hole'', with amplitude approximately  proportional to $M$,  
is apparently induced by the black hole at roughly its position during  the stalling phase.

\begin{figure*}
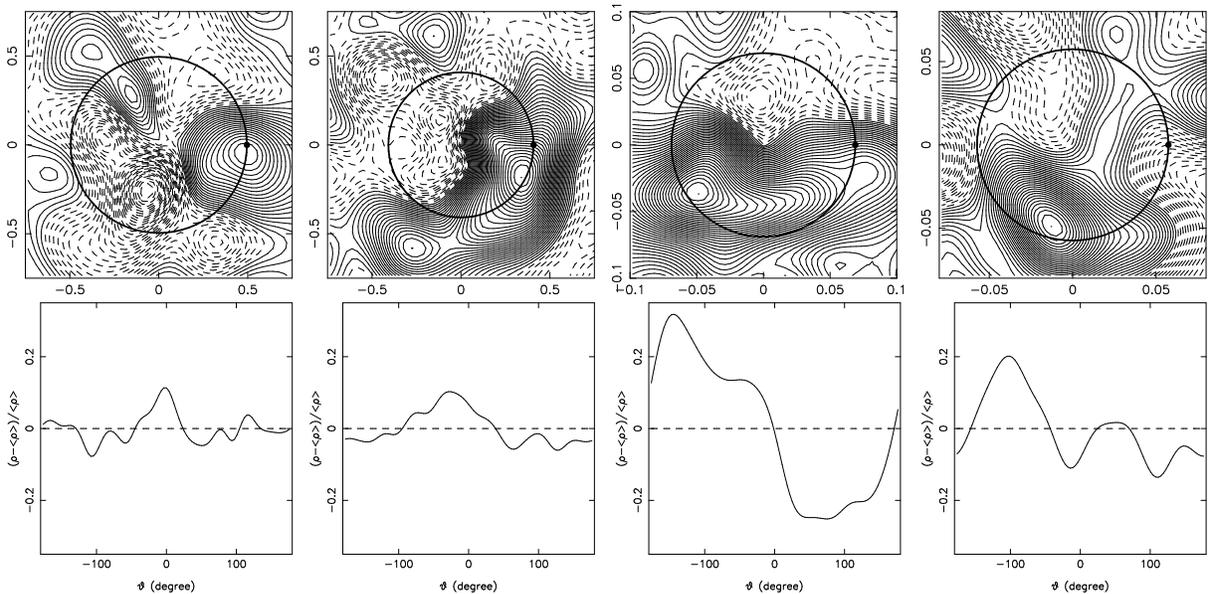

\begin{center}
$\begin{array}{cccc}  
 \includegraphics[angle=0,width=1.5in]{fig17a.eps} &    \includegraphics[angle=0,width=1.5in]{fig17b.eps}  &   \includegraphics[angle=0,width=1.535in]{fig17c.eps} 
 & \includegraphics[angle=0,width=1.5in]{fig17d.eps} \\
   \includegraphics[angle=0,width=1.5in]{fig17e.eps} &    \includegraphics[angle=0,width=1.5in]{fig17f.eps}  & 
     \includegraphics[angle=0,width=1.5in]{fig17g.eps}      &   \includegraphics[angle=0,width=1.5in]{fig17h.eps}                     \end{array}$           
  \caption{The density response (i.e., gravitational wake) induced by the massive body  
 in run E is shown  by plotting  density  contour maps of background stars in the upper
 panels, and the corresponding  relative overdensity along the black hole orbit in the bottom panels.
 The isodensity contours were obtained by subtracting at any radius the mean density and selecting only particles that were close to the orbital plane.
 Negative contours (underdensities)    are shown by dashed curves. Circular regions show the path over which the density was computed to obtain the plots in the bottom panels. }
\label{id}
\end{center} 
\end{figure*}

Figure~\ref{id} shows two-dimensional contour maps of the overdensity for
run E ($M=5\times10^4 $). The radial extension of the wake (with respect to
the galaxy center) does not change greatly over time,   but one can clearly see how the location of the density maximum shifts, and
  a density gap is apparently created near the black hole  position during the stalling phase. 

\begin{figure}{ }
\begin{center}  
\includegraphics[angle=0,width=2.4in]{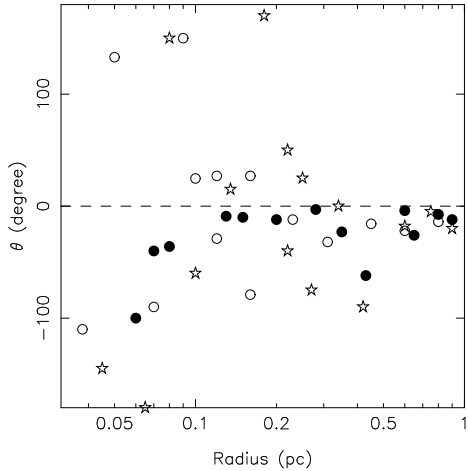}
  	\caption{ Position  of the relative density maximum as a function of the black hole galactocentric radius in runs A1 (filled circles), E (open circles) and F (stars symbols). 
  As in Figure~\ref{wake}, the $N$-body models were rotated such that the second black hole is  located at $\theta=0$ with $\dot{\theta}>0$. }\label{wake-pos}
  \end{center}
\end{figure}

To more clearly  illustrate  how the location of the gravitational wake with
respect to the second black hole evolves, we plot in Fig~\ref{wake-pos} 
the angular position of the maximum as a function of the black hole galactocentric radius. 
Outside the core (i.e., $r>0.3~$pc) the wake is located at small (negative) angles, causing 
the initial rapid inspiral.
Once the black hole starts to modify the background of stars the wake becomes more difficult to track. This causes the large oscillations seen in the 
relative position of the wake and in turn explains why such oscillations occur earlier for larger masses of the inspiraling object.

\section{Discussion} 

In this paper, we presented $N$-body simulations of  the inspiral of
a massive body into the Galactic center (GC).
Our models of the Milky Way nuclear star cluster were motivated by recent
observations that suggest a relatively low density of stars inside
the SMBH influence radius.
Such models are characterized by a zero or near-zero phase-space
density at low energies.
Under the standard approximation, in which the frictional  force 
from fast-moving stars is ignored,
a second black hole that sinks toward the center under the influence of 
dynamical friction would stall at a distance of roughly 1/2 the core radius, 
or $\sim 0.25$pc, from the SMBH. 
If the smaller black hole moves initially on a non-circular orbit,
its orbital eccentricity is predicted to increase with time due to the 
lower dynamical friction force near periapsis.

Using $N$-body simulations, we found that the frictional force never falls
precisely to zero.
As noted also by Chandrasekhar, stars moving faster than the test body 
contribute to the drag.
When this contribution is included in the expression for the dynamical
friction, Chandrasekhar's formula reproduces quite well the decay observed
in $N$-body simulations of the inspiral of a $\sim1000{M_{\odot}}$
black hole.
The eccentricity increase predicted by Chandrasekhar's theory is also confirmed.
When the inspiralling object is more massive,
a second mechanism contributes to the frictional force: the second black hole
induces evolution of  the background system, which tends to refill the initially
empty regions of phase space. 

In what follows, we discuss the implications of these results for a number of
astrophysical problems related to the dynamics of massive bodies near
the centers of galaxies.
But first, we comment on how our $N$-body results can be approximately
scaled to systems with different masses and densities.

The rate of inspiral of a massive body of mass $M$ is independent of the
mass of field stars if $M\gg m$.
Chandrasekhar's formula also predicts a linear dependence of the frictional
force on $M$\footnote{In its more general form (\ref{dfa}), 
the dynamical friction formula predicts
an additional, approximately logarithmic dependence of force on $M$.},  
and our simulations (as well as many others) confirm that prediction. 
If the density response of the background is ignored, the $N$-body results can then be scaled using:
\begin{eqnarray}
r &\rightarrow& r \times  \left[ \frac{ \tilde{r} \left(< 3.4 \times 10^{-3} M_{\bullet} \right)}{0.1~{\rm pc}} \right]; \\
 t &\rightarrow& t \times \left[ \frac{ \tilde{r} \left(< 3.4 \times 10^{-3} M_{\bullet} \right)}{0.1~{\rm pc}}   \right]^{3/2} \left[ \frac{M_{\bullet}}{4\times10^6 M_{\odot}}  \right]^{-1/2} \nonumber\\
&& \times \left[ \frac{\tilde{M}}{M}\times  \frac{4\times10^6 M_{\odot}}{M_{\bullet}} \right]^{-1},
\end{eqnarray}
where $\tilde{r}$ is radius containing  a mass in stars  $M_{\star}(<r) \approx 3.4 \times 10^{-3} M_{\bullet}$, $\tilde{M}$ is the mass of the test body 
and $M$ its mass adopted in the $N$-body simulations of  Table~1 .
When in the simulations the background of stars evolves, 
 the dependence on the mass of the infalling body becomes more complex;
in this case, the appropriate scaling is obtained by setting $\tilde{M}=M\times \left[  M_{\bullet}/4\times10^6 M_{\odot} \right]$, i.e., setting the ratio between the mass of the 
massive body and the central black hole the same as in the $N$-body simulation.
Particular caution should also be taken   when adopting  $\tilde{M}> M$ since for large values of $\tilde{M}$ the massive body would perturb the background system more importantly than
 it does in $N$-body runs.

The condition that the background not evolve is satisfied in our simulations
when $M\lesssim 10^4 {M_{\odot}}$ and at early times in run E.
We apply this approximate scaling to run A1, for which  $M= 5000{M_{\odot}}$
and the total integration time is $\sim 1.5\times 10^7 {\rm ~yr}$.
Assuming no change in the stellar density,
replacing the massive body by a $\sim 10~{M_{\odot}}$ black hole
increases the effective integration time by a factor $\sim500$, to $\sim 8\times10^9~{\rm yr}$ 
(at which time the galactocentric radius is  $\sim0.06~{\rm pc}$).
This result illustrates how -- in the absence of a steep central density cusp -- the time for 
stellar-mass BHs to reach the center of the Galaxy from a starting
radius of $\sim 1$ pc can easily exceed 
$\sim 10 {\rm Gyr}$ (a point we return to in Section~\ref{smr}).

Alternatively, we can identify our models with the center of a galaxy like M87, a luminous elliptical galaxy with a flat central density profile.
We adopt  $M_{\bullet}= 3\times 10^9~{M_{\odot}}$ for the 
mass of the SMBH and we use a core velocity dispersion  $\sigma_v =278 {\rm kms^{-1}}$ and the relation 
$\sigma_v ^2=4\pi G\rho_0(r_0/3)^2$ with $r_0=600~$pc to obtain the mass density profile \citep{yea,L:92}:
\begin{equation}\label{M87d}
\rho(r)=35~M_{\odot}{\rm pc^{-3}} \left( \frac{r}{600~{\rm pc}}\right)^{-\gamma}~.
\end{equation}
Taking $\gamma=0.6$, this  gives a length normalization factor $\tilde{r}\approx20~$pc.

\noindent Runs E and F: in these runs the background system evolves  due to the perturbations  
induced by the massive body (see Figure~14).  Setting  $\tilde{M}=M\times \left[  M_{\bullet}/4\times10^6 M_{\odot} \right]$, 
run E corresponds to the inspiral of a 
$\sim7\times10^6~{M_{\odot}}$ black hole
starting from a distance of $200~{\rm pc}$, and a total integration time $\sim 2 \times 10^9{\rm yr}$. 
In the case of run F, the inspiraling black hole would have a mass $\sim4\times 10^7~{M_{\odot}}$; it penetrates the inner $\sim 10~{\rm pc}$ in  $\sim3 \times10^8 {\rm ~yr}$ after which it effectively stalls.

\noindent Run A: the condition that the background not evolve is satisfied in runs A1, A2 and also at early times in run E. 
Setting $\tilde{M}=10^6 (10^5)10^4~{ M_{\odot}}$ in runs A1 and A2, the  final integration time and 
orbital radius are  $\sim 3 \times 10^8 (10^{10}) 10^{12}~{\rm yr}$ and $ 12~$pc respectively.  
This shows how, in  the central core of a M87-like galaxy, the inspiral time for black holes of masses $\lesssim 10^6~{ M_{\odot}}$
could easily exceed a Hubble time (a point we further discuss in Section~\ref{sbb}).

\begin{figure}{ }
\begin{center}  
\includegraphics[angle=270,width=3.3in]{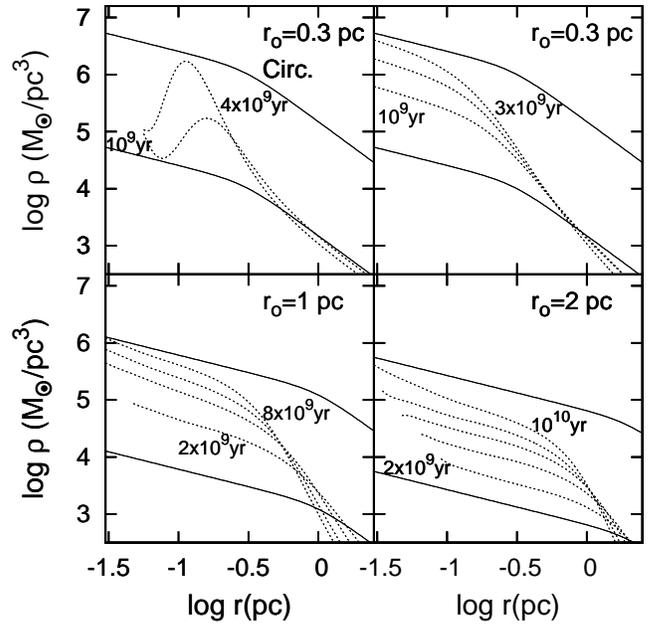}
  \caption{ Evolution of the density profile of a population of $10~{M_{\odot}}$ BHs (dotted curves) assuming that they constitute 
  $1\%$ of the total mass density initially. Results are displayed for three choices  of the core parameter $r_0=$(0.3, 1, 2)~pc.
   Lower (upper) solid lines show the initial density profile of stellar BHs (stars).
  In the upper-left panel the BHs lie on circular orbits while in the other cases we assume an isotropic initial distribution of velocities.
Density profiles are shown at time intervals of $\Delta t=2$x$10^9$Gyr in the lower panels, while $\Delta t=10^9$Gyr in the upper-right panel.  }\label{sbhs}
  \end{center}
\end{figure}

\begin{figure}{ }
\begin{center} 
\includegraphics[angle=270,width=3.3in]{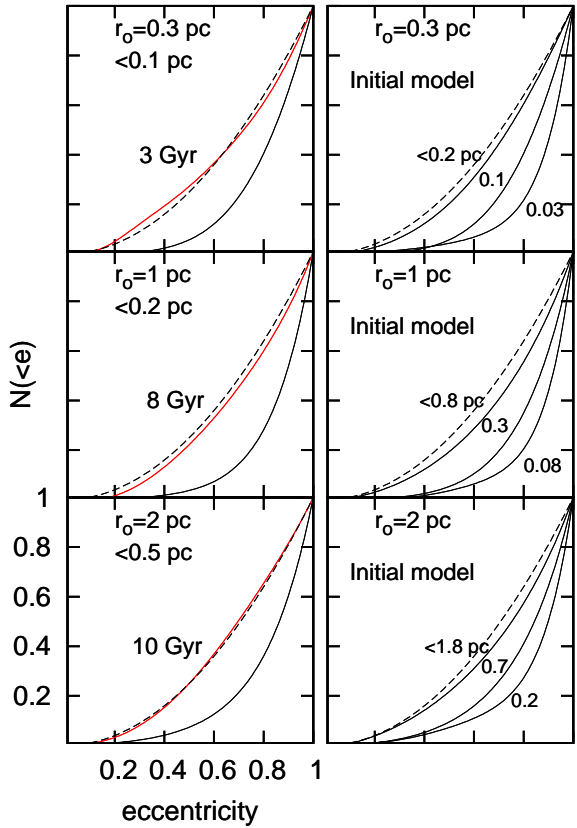} 
  \caption{ Left panels:  final cumulative eccentricity distribution of stellar BHs for the integrations displayed in Figure~\ref{sbhs} (red curves),
  that would be measured inside the core  within some  radius. Solid curves give the initial distributions.
   Right panels: cumulative eccentricity distributions of the initial models (solid curves) evaluated within different radii. At small galactocentric radii,
   the distribution is dominated by high eccentricity orbits, in spite of the fact that
  the velocity distribution is isotropic. Dashed curves show for comparison  a ``thermal'' eccentricity distribution, $N\sim e^2$. }\label{sbhs_ecc}
\end{center}
\end{figure}

\subsection{ Segregation of massive remnants at the Galactic center}\label{smr}
About $1\%$  of the total mass of the old population at the GC should be in the form of stellar-mass ($m\approx 10-20M_\odot$)  BHs \citep{AX:05}.
Since stellar BHs are significantly more massive than the mean stellar
mass ($\sim 1M_\odot$) expected for an evolved population, they would spiral 
in to the center and segregate around the SMBH \citep{M:93}.
The time evolution  of the remnant population  depends sensitively  on its initial
distribution and also on the  properties of the background distribution of lighter stars. 
In the case of a flat core in the stars, and a similar initial distribution in the BHs,
the  time for the latter to reach a steady state density profile can  exceed
a Hubble time, since the dynamical friction force essentially ceases inside the core
\citep{M:10}.
On the other hand, if the stars follow a steep central density cusp, 
the mass density of BHs after $\sim10~{\rm Gyr}$ can reach or  exceed that of the other populations 
within $\sim 10^{-2}~{\rm pc}$, leading to a quasi-steady-state density profile  
$n \simeq r^{-2}$ at small radii \citep[e.g.][]{HA:06,AH:09}.

Understanding the distribution of BHs at the centers of galaxies
like the Milky Way is crucial for making predictions about the expected event
rate for low-frequency gravitational wave detectors~\citep{HU:03}.
Since the stellar BHs at the GC are not directly detected,
time-dependent inspiral calculations like the ones presented here provide
the best hope of understanding their distribution.
However, if the background stellar distribution is a flat core,
our results show that a straightforward application of Chandrasekhar's 
formula can give misleading results.

Accordingly, we  computed the evolution  of a population  of stellar BHs 
as they spiralled in to the center of a galaxy with a flat stellar core, including the frictional
force from the fast-moving stars.
We began by generating random samples of positions and velocities from the  isotropic distribution function corresponding to the density
model of equation (\ref{den}) assuming $\gamma=0.6$; 
cores of various sizes, $r_0=(0.3, 1, 2)~{\rm pc}$;
and selecting only particles within $5~$pc of the SMBH.
In each of these models, a total of 800 orbits (representing the stellar BHs) were then integrated by solving the system of equations (\ref{em}), with dynamical friction force given by 
\begin{eqnarray}\label{form}
\boldsymbol{f_{\rm fr}} &=&-4\pi G^2  M \rho(r) \frac{\boldsymbol{v}}{v^3}  \Big( F(<v,r)  \ln\Lambda \\
 &&+\int^{\sqrt{-2 \phi(r)}}_v dv_\star 4\pi f(v_\star) v_\star^2    \left[ {\rm ln} \left( \frac{v_\star+v}{v_\star-v}  \right)-2\frac{v}{v_\star} \right]\Big) , \nonumber
\end{eqnarray}
with $\ln\Lambda=15$, $M=10~{M_{\odot}}$.
At each time,  the density profile and eccentricity distribution of the inspiralling
objects were computed by sampling each orbit 
over time intervals of $0.3~{\rm Gyr}$.
We also considered one model with  core parameter $r_0=0.3~{\rm pc}$ in which  
all BHs were initially on circular orbits.

All of the calculations presented in this section
assume that the mass density due to the BHs 
remains small compared with the mass density in stars,
and that the stellar distribution is unchanging.
Because the two-body relaxation  time for $1~{M_{\odot}}$ stars is so long
in these models,  and $\sim 10$ times longer than the black hole inspiral time,
ignoring the evolution of the stellar distribution due to star-star encounters is 
reasonable.  This basic assumption is also supported by  recently published $N$-body                                                           
 simulations~\citep{GM:11} that show how, in models with a pre-existing stellar core,                                           
 the distribution of BHs evolves against an essentially fixed background of stars.
However, once the density in BHs begins to approach that in 
the stars, our calculations are no longer valid.

In Figure \ref{sbhs} we plot the  density profile of BHs at different times, assuming that their fraction is initially $10^{-2}$
of the total mass density.  The upper panels give the results for the model with  $r_0=0.3~$pc. In these integrations the core is very small
and  after only $\sim 1~$Gyr the density  of black hole rises  very steeply going into the stellar core.
After $\sim 4~$Gyr the BHs accumulate  at  radii near the core,
matching the density in stars at $\sim 0.01~$pc. 
In the circular-orbit model, the density profile 
at $1~$Gyr shows a maximum at $\sim 0.2~$pc, that grows and migrate inward due 
to the friction produced by fast moving stars inside these radii.
The evolution for  the isotropic run is comparably rapid, and after $\sim 3~$Gyr the density of BHs reaches that in stars at $\sim 0.01~$pc.

\citet{M:10} showed that a core of the size currently observed is a natural consequence
of two-body relaxation acting over $10~$Gyr, starting from a core of radius $\sim1~$pc.
It is therefore of interest to study the evolution of the black hole distribution in models with parsec-scale cores.
This is shown in the lower panels of Figure \ref{sbhs}.
In these cases the evolution is slower as a consequence of the increased size of the stellar core, and even after $10~$Gyr the density of BHs can remain  substantially lower than that in stars at all radii.
We conclude that  it would be unjustified to assume
that the massive remnants have yet reached a steady-state density  profile at the GC.
One consequence  is that rates of capture of stellar 
BHs by the supermassive black hole at the Galactic center (EMRIs) may be much lower than 
in standard models that postulate a collisionally-relaxed nucleus~(e.g., Hopman \& Alexander~2006).

The left panels of Figure \ref{sbhs_ecc} plot  the cumulative distribution of eccentricities of BHs inside various radii.
Since the final eccentricity of each orbit is larger than its initial value (see Section 2.2), one might naively expect the eccentricity distributions to evolve toward
a form that is increasingly strongly peaked near $e\approx 1$.
This would be the case if one plotted $N(e)$ for a {\it fixed} subset of objects.
However, when restricting the sample to a given {\it radial} range,
the result is very different.
The reason \cite[e.g.][Appendix]{M:10} is illustrated in the right-hand panels
of Figure~\ref{sbhs_ecc}: given a flat density profile, even an isotropic distribution 
of objects around a SMBH will have  an eccentricity distribution that is strongly
peaked near $e=1$, since the only objects that can approach closely to  the SMBH
are on highly eccentric orbits.
As the distribution of BHs evolves away from this initial configuration,
the regions of low-energy phase space that were initially empty are gradually
refilled, and the eccentricity distribution begins to approach more closely
to a ``thermal'' form, $N(<e)\propto e^2$.
In addition, (i) the eccentricity of individual orbits inside the core 
grows only very slowly  since they are in a region where the dynamical friction 
force is small (see Figure 3); 
(ii) the eccentricity of BHs initially beyond the core decreases 
initially since they lie in a $\gamma\approx 1.8$ cusp; their eccentricities
subsequently  increase as the orbital periapsis  enters the core, but in most cases   
this second phase is   too short (see Figure~$~4$) to produce final  eccentricities significantly different from the initial values.
We finally computed the anisotropy parameter (\ref{eq:beta}) 
at the final integration time, defined as the time when the mass density in BHs reaches that in stars at small radii, and
found that the departures from isotropy remained small at all radii.

\begin{figure*}
\begin{center} 
$\begin{array}{c} 
\includegraphics[angle=270,width=6.2in]{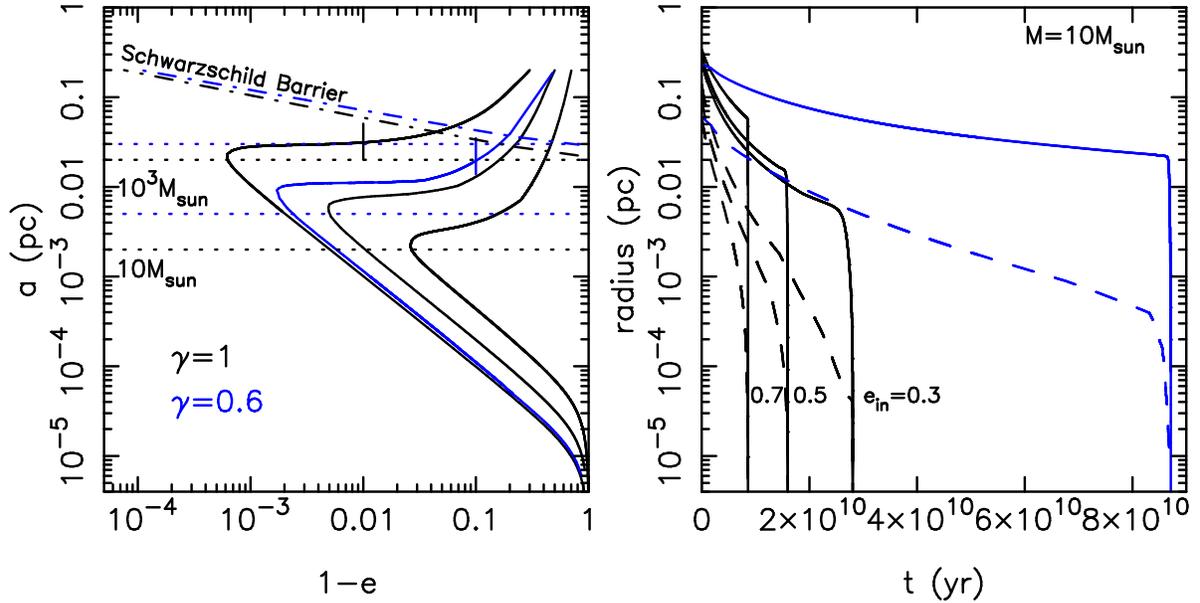} 
\end{array} $
\caption{Left panel: evolutionary tracks  of a massive object in the Galactic center starting from various eccentricities $e_{\rm in}=(0.3, 0.5, 0.7)$,
 from an initial semi-major axis $a_{\rm in}=0.2~$pc and adopting two different inner slopes of the  mass-density profile 
$\gamma=(1, 0.6$).  Dot-dashed lines are the Schwarzschild barrier, equation (\ref{sb}), below which  resonant relaxation
is suppressed by relativistic precession and gravitational scattering is dominated by classical non-resonant relaxation.
Vertical marks give the radii within which the two-body relaxation time scale for changes in angular momentum ($t_{r,\rm eff}$) becomes shorter than 
the time-scale of evolution for angular momentum  in our integrations ($t_{\rm evol}$), assuming $10~{M_{\odot}}$ for the mass of the inspiraling black hole.
Inside these radii,  for  $M\leq10~{M_{\odot}}$, our integrations are no longer valid since two-body scattering, rather than dynamical friction, would dominate the
orbital evolution.
For $e_{\rm in} \lesssim 0.5$ and $\gamma=1$ (two right-most curves), at any radius, $t_{r,\rm eff}$ was always longer than $t_{\rm evol}$  and no vertical marks are displayed.
In the two left-most curves, the condition that $t_{r,\rm eff} > t_{\rm evol}$ at any radius wold instead require a slightly larger mass 
for the BH: $M\gtrsim 15~M_{\odot}$.
This shows that  gravitational scattering from stars can be neglected and our integrations are valid for relatively small masses of the test particle.
Within the Schwarzschild barrier,  dynamical friction is therefore the main mechanism inducing  creation  of EMRIs.
We also stress that in these integrations, changes in the stellar distribution are not taken into account.
For instance, the stellar potential would be strongly perturbed when  the
mass of the inspiraling black hole  becomes comparable to 
 the mass in stars contained inside its orbital radius. 
As a reference, dotted lines in the panel display the radius within which the mass in stars in the model is $10$ or $1000~M_{\odot}$.
Right panel:  time evolution of  periapsis (dashed lines) and apoapsis  (continue line)  for a $10~{M_{\odot}}$  BH. 
The sinking time scale decreases with increasing the initial eccentricity,
 and, for the set of computed orbits, it is shorter than $10^{10}~$yr  only for $e_{\rm in}=0.7$  (left-most curve in the panel).
}\label{f+r}
\end{center}
\end{figure*}

\begin{figure}
\begin{center} 
$\begin{array}{c} 
\includegraphics[angle=270,width=3.2in]{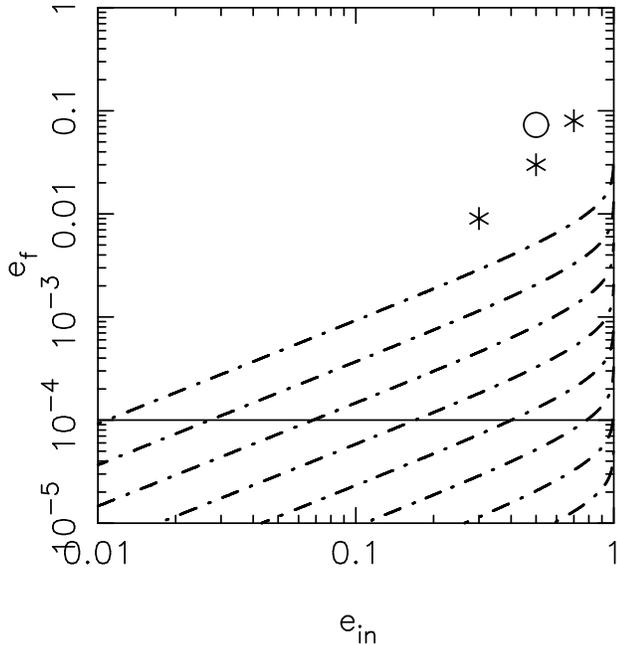} 
\end{array} $
\caption{Eccentricity at the moment the binary enters the sensitivity window of  planned  space-based interferometers, $e_f$,
as a function of the initial orbital eccentricity $e_{in}$  for the
integrations displayed in Figure~\ref{f+r}.  Star symbols are for $\gamma=1$ (black curves in Figure~\ref{f+r}),
empty circle  for  $\gamma=0.6$  (blue curve in Figure~\ref{f+r}).
 The dot-dashed lines  give $e_f$ ignoring dynamical friction. 
For a given initial eccentricity and secondary black hole mass  we  fixed the merger time  by using equation~(\ref{tmer}) and  varying the initial orbital semi-major axis.
If we take a test mass of $10~M_{\odot}$ ($1000~M_{\odot}$) this corresponds  to merger times of $10^{15},~10^{14},... 10^{8}~$yr 
($10^{13},~10^{12},... 10^{6}~$yr) from  bottom to  top line. 
As comparison, the orbital eccentricity and merger time for the integrations of Figure \ref{f+r}, at the moment  GW energy loss stars to dominate the evolution,
are (from left to right of that figure) $e\sim (0.9994, 0.998, 0.994, 0.97$) and $t_m\sim(5.9\times10^7,~1\times10^8,1.9\times10^9,~5\times10^9)~$yr. 
Horizontal line represents
approximately the  lowest value of $e_f$ that would require non circular  templates for data analysis~($e\sim 10^{-4}$, Porter \& Sesana~2010).
}\label{fr+gr2}
\end{center}
\end{figure}

\subsection{Dynamical Evolution of Eccentric  Black Hole Binaries} \label{dfde}
Gravitational radiation emitted by binary  black holes   with masses $10^3-10^7M_{\odot}$
 is the principal target  of  planned, space-based,  gravitational wave  observatories.
In the present literature the strain amplitude of the gravitational wave (GW) radiation is typically obtained under the
assumption of  complete circularization of the binary  at the moment that the signal 
enters into the observable  band.
This simplification is  motivated by the predicted strong
eccentricity  decay   when binary dynamics  are
dominated by relativistic effects.
The expressions of the time average  change of 
 eccentricity  $e$ and  semi-major axis $a$  in the relativistic regime of a binary with components of
masses $m_{\rm 1}$ and  $m_{\rm 2}$  were derived  by Peters \& Matthews~(1963):
\begin{eqnarray}
  \left\langle\!\frac{da}{dt}\!\right\rangle  =
 -\frac{64}{5} \frac{G^3\, m_1\, m_2 (m_1+m_2)}{c^5\, a^3} f(e), ~~~~~~\\
  \left\langle\!\frac{de}{dt}\!\right\rangle  =
  -\frac{304}{15} \frac{G^3\, m_1\, m_2 (m_1+m_2)}{c^5\,a^4
  (1-e^2)^{5/2}} \left(e+\frac{121}{304}e^3\right),  
\label{dedtpm}
\end{eqnarray}
where $c$ is the speed of light and   
 \begin{equation}
f(e)=\left(1-e^{2}\right)^{-7/2}\left(1+\frac{73}{24}e^{2}+\frac{37}{96}e^{4}\right)  \label{enh}~~.
\end{equation}
The strong dependence of the enhancement factor $f(e)$  on $e$,
shows the  fundamental role of the binary eccentricity in determining the
rate at which the system loses energy due to   GW emission.

A way to follow the  orbital inspiral of a massive body at the GC, due both to dynamical friction and 
gravitational wave radiation,  is to couple  Chandrasekhar's formula   for the frictional drag
with the 2.5 post-Newtonian equations  representing  GW energy loss~\citep{Merritt:12}. 
In the limit $M / M_{\bullet}<<1$, the total deceleration can be approximated by
\begin{eqnarray} \label{df+gr}
\boldsymbol{f} &=&- 4\pi G^2  M \rho(r) \frac{\boldsymbol{v}}{v^3}  \Big( F(<v,r)  \ln\Lambda \\
&&+ \int^{\sqrt{-2 \phi(r)}}_v dv_\star 4\pi f(v_\star) v_\star^2    \left[ {\rm ln} \left( \frac{v_\star+v}{v_\star-v} \right)-2\frac{v}{v_\star} \right]\Big) \nonumber\\
&& -GM\left [A \boldsymbol{n} + B \boldsymbol{v} \right] \nonumber
\end{eqnarray}
where $\boldsymbol{n}=\boldsymbol{r}/r$ and
\begin{eqnarray}
A=\frac{1}{c^5}\left[ -\frac{24 v_r v^2 G M_{\bullet}}{5~r^3} - \frac{136 v_r G^2 {M_\bullet^2}}{15~r^4} \right]~; \\
~~B=\frac{1}{c^5}\left[ \frac{8 v^2  G  M_{\bullet}}{5~r^3} + \frac{24 G^2 M_{\bullet}^2}{5~r^4} \right]~,~~~~~~~
\end{eqnarray} 
with  $v_r$ the radial component of the velocity vector. 
Evidently, both the frictional force and the 2.5PN correction are dissipative terms, but, while the latter term always drives to lower eccentricities,
the effect of dynamical friction on the orbital eccentricity  has a strong dependence 
on  the phase-space distribution associated with the stellar background  (see \S~\ref{ecc}).
 Notice that, in  equation  (\ref{df+gr}),
  if we neglect the dependence of the Coulomb logarithm on the mass of the test body, both the frictional term 
and the post-Newtonian terms depend linearly on $M$  implying that the  time evolution of the orbital 
elements  can be trivially rescaled to any  $M$ as long as the condition $M / M_{\bullet}<<1$ holds (see also equations 36 and 37).

\subsubsection{Dynamical Friction in the Context of the EMRI Problem}
Extreme-mass-ratio inspirals (EMRIs) are a potential source of low-frequency   gravitational waves \citep{HU:03,BC:04,AS07}.
In steady-state models of the Galactic center, the distributed mass within $10^{-2}~$pc of the SMBH is dominated by stellar BHs~\citep{HA:06}.
At these radii, dynamical friction is therefore typically ignored and it is assumed that captures for EMRIs are driven by 
gravitational scattering   from other stellar BHs \citep[e.g.][]{MAMW:11}.
On the other hand, if the  background stellar distribution  has a flat core, the  density of  BHs  can remain 	
small compared with the mass density of other populations 
\citep[e.g.][\S6.1]{GM:11}. Under these circumstances,
 at any radius, massive remnants might see a background  whose density comes mostly from lighter stars and dynamical friction  becomes a competing  mechanism 
in driving capture of EMRIs.

Using equation (\ref{df+gr}) we computed the trajectory  of the test mass under a variety of assumptions  for the  background system. 
Results of these integrations are displayed in Figure~\ref{f+r}.
We considered orbits of different initial  eccentricities $e_{\rm in}=(0.3, 0.5, 0.7)$,  starting from a semi-major axis $a_{\rm in}=0.2~$pc. 
For the stellar background  we used the density model of equation~(14)  with two different values of the internal slope index: $\gamma=1$ (black lines) and 0.6~(blue lines). 
 For  an eccentric orbit in  a flattened  cusp, dynamical friction at apoapsis dominates the evolution causing a rapid 
increase of the orbital eccentricity. In the simplified  picture in which the frictional force at periapsis is vanishing small, the apoapsis distance remains unchanged in time,
while the periapsis becomes progressively smaller;  at some point, the  minimum distance from the 
SMBH is small enough  that the 2.5PN terms start to dominate the evolution.
The drag at periapsis then circularizes the orbit, and  causes the merger of the two black holes.

Near a SMBH, as long as the relativistic  precession time scale is much longer than the orbital period, the mechanism that dominates the scattering of 
stars onto high-eccentricity orbits is resonant relaxation. Because in the potential of a point-mass the orbits are fixed ellipses, perturbations on a test particle 
are not random but correlated \citep{RT:96}.
The residual torque $|{\boldsymbol T}|\approx \sqrt{N}Gm/r$, exerted by the $N$ randomly oriented, 
orbit-averaged mass distributions of the surrounding stars, induces   coherent changes in angular momentum $\Delta {\boldsymbol L}={\boldsymbol T}t$ on times $t\lesssim t_{coh}$,
where the coherence time $ t_{coh}$ is fixed by the mechanism that most rapidly causes the orbits to precess (e.g, mass precession, relativistic precession).
The angular momentum relaxation time associated with resonant relaxation is:
\begin{equation}\label{rs}
t_{rr} = \left( \frac{L_c}{\Delta L_{coh}} \right)^2 t_{coh}~,
\end{equation}
where $L_c \equiv \sqrt{G M_{\bullet} a} $ is the angular momentum of a circular orbit and $|\Delta L_{coh}|\sim |{\boldsymbol T}t_{coh}|$ is the 
accumulated change over the coherence time.
Assuming that the precession is determined by the mean field of  stars, the angular momentum relaxation time becomes \citep{RT:96}:
\begin{equation}\label{rs2}
t_{rr}\approx 2.9 \times 10^7  {\rm yr}\left( \frac{M_{\bullet}}{4\times10^6{M_{\odot}}}\right)^{1/2} \left( \frac{a}{{\rm 0.1 pc}} \right)^{3/2}  \left( \frac{m}{ {M_{\odot}}} \right)^{-1}~.
\end{equation}

Dot-dashed lines in the left panel of Figure~\ref{f+r} give the Schwarzschild barrier.
Above these lines,  resonant relaxation is the most rapid mechanism affecting angular momenta;
below the curves, relativistic precession becomes  efficient at
suppressing resonant relaxation and the gravitational perturbations are dominated by  classical "two-body" relaxation.
The value of the angular momentum 
that defines the Schwarzschild barrier is \citep{MAMW:11}:
\begin{eqnarray}\label{sb}
(1-e^2)_{\rm SB}&\approx& 5.8\times 10^{-3} \left( \frac{C_{\rm SB}}{0.7} \right)^2 \left( \frac{a}{{\rm 0.1~pc}} \right)^{-2}  \\
&&\times \left( \frac{M_{\bullet}}{4\times10^6{M_{\odot}}} \right)^{4}
\left( \frac{m}{ {M_{\odot}}} \right)^{-2} \left( \frac{N}{10^4} \right)^{-1}~,\nonumber
\end{eqnarray}
where $N$ is the number of stars within radius $a$ and  $C_{\rm SB}$ is a constant of order of unity.
Beyond the barrier, the time for encounters to change the orbital angular momentum by of order itself is
$t_{r,\rm eff}=2(1-e)t_{\rm r}$, where  for the non-resonant relaxation time scale  we adopt the approximate expression \citep{HA:06}:
\begin{eqnarray}\label{tr2}
t_{r} &\approx& 4.8 \times 10^{10} {\rm yr} \left( \frac{a}{{\rm 0.1~pc}} \right)^{3/2} \left( \frac{M_{\bullet}}{4\times 10^6 {M_{\odot}}} \right)^{3/2} \\
&&\times \left( \frac{m}{ {M_{\odot}}} \right)^{-2} \left( \frac{N}{10^4} \right)^{-1}~. \nonumber
\end{eqnarray}
For our integrations to be viable, the time-scale for dynamical friction to change the orbital angular momentum, $t_{\rm evol} \sim (1-e)|{\rm d}(1-e)/{\rm d}t|^{-1}$, 
must be shorter than $t_{r,\rm eff}$ at all radii.
Vertical  marks in the left panel of Figure~\ref{f+r}  give the orbital radius within which $t_{r, \rm eff}$ becomes smaller than $t_{\rm evol}$  
assuming  $M=10~M_{\odot}$.
For $e_{\rm in}=(0.3,~0.5)$ and $\gamma=1$ (two right-most curves), at any radius, $t_{r,\rm eff}$ was always larger than $t_{\rm evol}$  and no vertical marks are displayed.
Increasing the initial eccentricity to $e_{\rm in}=0.7$, $t_{r, \rm eff}$ equals  $t_{\rm evol}$ at $\sim 0.03~$pc. At smaller radii,
  two-body relaxation would dominate the orbital evolution and, for a $10~M_{\odot}$ black hole, this integration is not longer valid. 
Taking $e_{\rm in}=0.5$ and  $\gamma=0.6$ (blue curve in the figure) this transition occurs at  $\sim 0.02~$pc.
Because increasing the mass of the test body reduces $t_{\rm evol}$ but leaves $t_{r,\rm eff}$ unchanged, it is always possible to set $M$ such that
the condition $t_{r,\rm eff}>t_{\rm evol}$ is satisfied  everywhere  within the Schwarzschild barrier.
In these two latter cases  this condition requires a slightly larger mass of the test body: $M \gtrsim15~M_{\odot}$~\citep{WHW}.
Two-body scattering effects from field stars can therefore be ignored for relatively small masses of the sinking black hole.
We conclude that, in a flat density distribution near a SMBH and at radii relevant for the EMRI problem ($<0.01~$pc),  dynamical friction  might be 
an important process in driving the formation of EMRIs.

Gravitational scattering  can be dominated by  other stellar black holes if their density becomes  comparable of that in stars at small radii as a consequence of mass-segregation.
In an unsegregated model the number of stellar black holes (of mass $10~M_{\odot}$) is predicted to be $10^{-3}$ times that in stars. From equations~(\ref{sb}) and (\ref{tr2}) 
it follows that,  in this case, the scattering from black holes  can be ignored with respect to the perturbations from the
stellar population. 
Gravitational scattering from black holes starts to compete with  that from stars  when their number  at small radii ($\sim 1~$mpc)
 is $10^{-2}\times N$, similar to the found at later times in Figure~\ref{sbhs} for $r_0\sim2$pc.
In relaxed mass-segregated models, instead, the number of black holes would be  approximately  $N$, and they will 
dominate the orbital evolution of the test mass at any radius inside the Schwarzschild barrier~\citep{AH:09}.

Finally, we note that dynamical friction can be very  inefficient if the mass of the inspiraling object  becomes  comparable  to the mass in stars within its orbital radius.
In the $\gamma =1$ cusp for a $\sim 1000(10)~{M_{\odot}}$, this occurs  at $\sim 0.02(0.002)~$pc
or at  $\sim 0.03(0.005)~$pc when  $\gamma =0.6$. 
This suggests that the results of Figure~\ref{f+r}  may not apply for large masses of the test body  and
for small initial eccentricities ($\lesssim 0.3$).
Accurate $N$-body simulations, including high-order  post-Newtonian terms, 
should be used  to better understand at which extends the conclusions made here  can be applied. We reserve this study to a future paper.

In order for an extra-galactic source to be observable by   proposed space-based interferometers, 
 it must have an orbital frequency $\gtrsim 10^{-4} {\rm Hz}$ \citep{AS07}, or 
\begin{equation}\label{la}
a \lesssim a_f \equiv 4 \times 10^{-3} {\rm mpc} \left( {\frac{ M_{\bullet}}{4 \times 10^6 {M_{\odot}}}} \right)^{1/3}~.
\end{equation} 
We explored  whether the computed orbits would retain some degree of eccentricity by the time the binary 
enters  the instrumental  sensitivity window, 
by evaluating   the eccentricity, $e_f$, at the time at which the condition~({\ref{la}}) is 
satisfied and comparing  this value  with the  minimum eccentricity that would require non-circular templates for data analysis: $e\sim 10^{-4}$ (Porter \& Sesana~2010). 
We note that strong sources (with high eccentricity) might be detectable at lower frequencies (i.e., larger semi-major axis)  \citep{AS07}.
The use of equation~(\ref{la}) is therefore a conservative one.

Figure~\ref{fr+gr2} plots $e_f$  as a function of the initial eccentricity 
for the orbits displayed in Figure~\ref{f+r}. 
In addition,  we computed a set of orbits with different initial eccentricities by removing from equation~(\ref{df+gr}) the dynamical friction term.
Each dot-dashed curve in the figure corresponds to a fixed value for the coalescence time~\citep{pet64}:
\begin{eqnarray}  \label{tmer}
t_m &\simeq&   3.6\times10^{12} {\rm yr} \left( \frac{10M_{\odot}}{M} \right) \left( \frac{4 \times10^6 M_{\odot}}{M_{\bullet}} \right)^2  \\
&&\times \left( \frac{a}{\rm mpc} \right)^4(1-e^2)^{7/2}~. \nonumber
\end{eqnarray}
Taking $M=10~M_{\odot}$ ($1000~M_{\odot}$), this corresponds  to  $t_m=10^{15},~10^{14},... 10^{8}~$yr 
($10^{13},~10^{12},... 10^{6}~$yr) from the bottom to the top line  respectively. 
It is evident that even for relatively low initial eccentricities and large merger times the binary will have a value of $e_f$ significantly different from zero.
This study suggests that secondary black holes typically  reach the GW radiation regime on wide orbits that are still very eccentric,
and even after the semi-major axis has decreased to values small enough for detection by space-based interferometers, eccentricities can be large enough that the efficient analysis of gravitational wave signals would require the use of eccentric templates  \citep[see aslo ][]{BC:04}.

\begin{figure*}
\begin{center}
$\begin{array}{c}
\includegraphics[angle=270,width=6.8in]{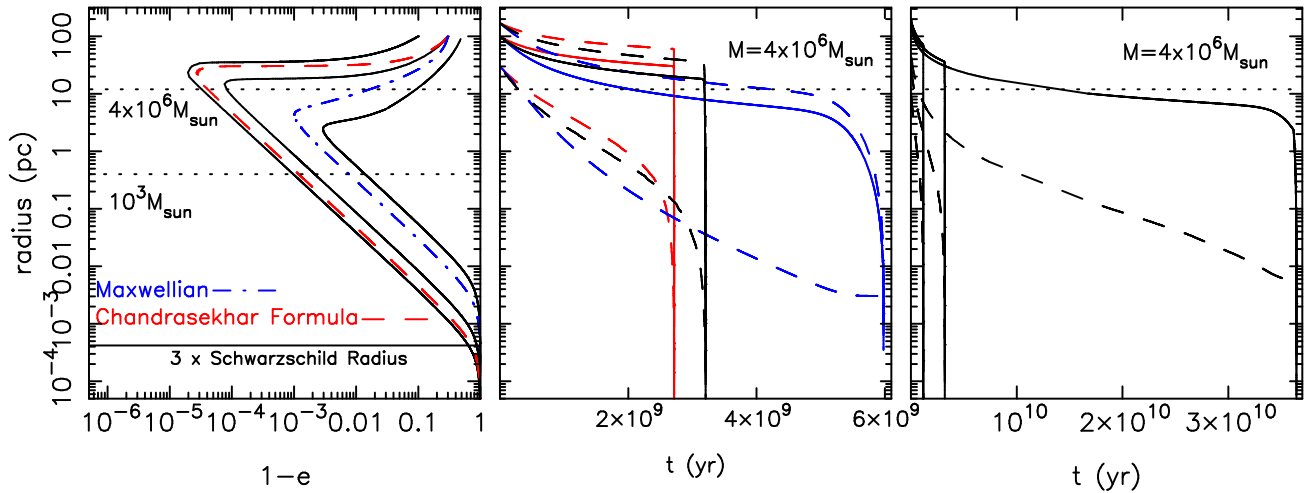}
\end{array} $
\caption{Left panel: orbital evolution in the $a, (1-e)$ plane for a massive object in the M87 core starting from various eccentricities $e_{\rm in}=(0.5, 0.7, 0.9)$,
and from an initial semi-major axis $a_{\rm in}=100~$pc.
Dynamical friction and gravitational-wave energy losses are both included.
Dotted lines represent the radii at which the stellar mass enclosed
in the orbit is $10^3~{M_{\odot}}$ (lower curve) or $4\times10^6~{M_{\odot}}$ (upper curve).
Red and blue lines are obtained respectively from the standard Chandrasekhar formula~(\ref{dfa}), which neglects fast-moving stars, and from equation~(\ref{ff})
that assumes in addition a Maxwellian velocity distribution.
Black curves are based on the more general equation~(\ref{df+gr}).
Horizontal solid line gives the ISCO radius for a non-spinning hole (i.e., 6 gravitational radii).
Central panel: time evolution of orbital semi-major axis (solid lines), apoapsis (upper-dashed curves) and periapsis (lower-dashed curves) in
the three integrations with $e_{\rm in}=0.7$ performed using: (i) the correct formula that includes the contribution from fast moving stars (black curves),
(ii) equation~(\ref{dfa}) in which only stars moving slower than the test mass contribute to the frictional drag (red curves); (iii) equation~(\ref{ff}) which assumes a Maxewllian distribution of velocities (blue curves).
Right panel: time evolution of apoapsis (solid lines) and orbital periapsis (dashed lines) for a $4\times10^6~{M_{\odot}}$
black hole.\label{M87}}
\end{center}
\end{figure*}

\subsubsection{Orbital decay in the cores of giant elliptical galaxies}\label{sbb}
Until the discovery of a stellar core in the Milky Way  \citep{BSE,D:09,B:10},
the density was generally assumed to follow a steep power law,
$\rho\sim r^{-2}$, inside the influence radius of Sgr A$^*$.
The same assumption is still commonly made when modelling the so-called
``power-law'' galaxies: galaxies of low to moderate luminosity that also
exhibit steeply-rising densities near the center \citep{GEB:96,FAB:97}.
Whether other power-law galaxies will turn out to harbor parsec-scale 
cores like the one in the Milky Way remains to be seen.
But it has long been known that cores are ubiquitous in
stellar spheroids brighter than $\sim10^{10}L_\odot$, whose influence
radii can be resolved \citep{F:94,L:95}.
Core sizes are observed to be of order the SMBH influence radius or somewhat
greater, consistent with models in which the cores are produced by
the scouring effect of binary SMBHs \citep{MM:06,GM:11}.

In this section, we use equation~(\ref{df+gr}) to investigate the orbital evolution of
a massive black hole that spirals in to the center of a giant elliptical galaxy with a core.
We base our models on M87.
The relevant properties of M87 are summarized at the start of this section.
Here we note that the core of M87
extends substantially beyond the SMBH influence radius:
$r_c/r_\mathrm{bh}\approx 600 \mathrm{pc}/200 \mathrm{pc}\approx 3$.
By comparison, the Milky Way has $r_c\approx0.3 r_\mathrm{bh}$.
This difference may reflect different formation processes for the two cores,
or may be a result of the  shorter relaxation time at the center of the
Milky Way, which could cause the core to shrink over 10 Gyr \citep{M:10}.

Following the evolution of a binary SMBH at the center of a galaxy requires self-consistent
simulations that can correctly treat the response of the background stars to the 
presence of the second massive body
\citep[e.g.][]{EMO,QH97,MM}.
Here, we limit ourselves to the case where the inspiralling black hole
is much less massive than the central one.
For instance, capture of a Milky-Way-sized galaxy by M87 would bring a second
SMBH into the center forming a binary of mass ratio $\sim 10^{-3}$.
This problem can be seen as a scaled-down version of the capture of an intermediate-mass
black hole by Sgr A$^*$.
Simulations of the latter scenario  \citep[e.g.][]{BGP:06} have generally assumed a steeply-rising
stellar density around the SMBH; inspiral of the intermediate mass black hole  is found to stall when
the semi-major axis of the binary drops to $\sim 10^{-3}$ pc, the radius at which the
binary is able to eject stars with greater than escape velocity.
When there is a pre-existing core, the binary evolves somewhat differently than
in these simulations; as we showed above,
the orbital periapsis progressively decreases while the apoapsis hardly changes.
As a result, the orbital semi-major axis can still be large at the time that GW losses becomes significant.
Since most of the frictional force occurs near apoapsis, we do not expect significant
stalling or core depletion to occur until late in the evolution, perhaps not before the two black holes merge~\citep[e.g.][]{FEM:92}.
Nevertheless, in what follows, we will explicitly note when in our integrations the mass of the sinking object becomes comparable to the mass in stars enclosed within its orbital radius.

We carried out calculations using the mass density profile of equation~(\ref{den})
with~\{$\alpha=1; \gamma_e=1.8; \gamma=0.5; r_0=600~{\rm pc}; \rho_0=35~M_{\odot} {\rm pc^{-3}}$\}, and $M_{\bullet}=3\times 10^9~M_{\odot}$ .
The left panel of Figure~\ref{M87} gives the orbital evolution of a test particle starting from an orbital radius of $100~$pc, and eccentricity
$e_{\rm in}=(0.5, 0.7, 0.9)$.
Dotted lines in the panel represent the radii at which the stellar mass enclosed
in the orbit is $10^3$ or $4\times10^6~{M_{\odot}}$.
For the two more eccentric orbits (two left-most curves),
it is possible that the binary enters the GW regime before violating these conditions.

Although in our model the binary black hole mass  is above the range  ($10^3-10^7~M_{\odot}$) normally associated with
space-based interferometers, we can nevertheless ask
whether the eccentricity would remain large after the massive binary 
has entered into the GW regime.
The Schwarzschild radius of a $3\times10^9~M_{\odot}$ SMBH is $r_{\rm SC} \approx 1.4 \times 10^{-4} {\rm pc}$.
When the orbital semi-major axis is $10\times r_{\rm SC}$, we find that the binary eccentricity is still very large: $e\approx (0.08,~0.6,~0.8)$
for $e_{\rm in}= (0.5,~0.7,~0.9)$.
When $a=5\times r_{\rm SC}$ the corresponding eccentricity is $e\approx (0.03,~0.4,~0.7)$.

The blue curves in Figure~\ref{M87} were obtained by computing the frictional drag using
Chandrasekhar's formula in its most common form, which assumes a locally
Maxwellian distribution of velocities (equation~[\ref{ff}]).
This approximation results in a very different orbital evolution characterized by
smaller orbital eccentricities (for a given semi-major axis) and faster orbital decay
when compared with the results obtained using the more correct formula~ (\ref{df+gr}).
We note that -- in spite of a higher rate of orbital decay -- the smaller eccentricities achieved during the infall in this case result in a longer lifetime of the massive binary (central panel).
The red curves in Figure~\ref{M87} were obtained using equation~(\ref{dfa}),
which allows for a non-Maxwellian velocity distribution but neglects the contribution
to the frictional drag from stars moving faster than the sinking black hole.
This approximation also results in a very different evolution when compared to the
more correct treatment (black curve).
Due to the smaller frictional drag, the standard treatment produces a slower decay of the orbital semi-major axis
but a much faster evolution of the eccentricity which in turns results in a shorter life-time of the black hole binary.
The right panel of Figure~\ref{M87} shows the time evolution of orbital radius when $M=4\times10^6~{M_{\odot}}$.
In a shallow cusp near a SMBH, dynamical friction is very inefficient; this results in a very long sinking time.
Starting from $100~$pc, black holes with masses $M \lesssim 4\times10^6~{M_{\odot}}$
do not reach the center of the galaxy in a Hubble time unless their orbit has a substantial
initial eccentricity ($e_{\rm in}\lap 0.7$).
We note however that such large eccentricities could be difficult to retain at these radii due to orbital 
circularization that occurs outside the sphere of influence of the central SMBH.

Cosmological simulations predict that a giant elliptical like M87  accreted  about 
 $4$ Milky-Way-sized galaxies over the last $\sim 5~$Gyr \citep{FMB}.
 The long sinking time scales found  in Figure~\ref{M87}  suggest therefore
 that, at the present epoch,  brightest cluster galaxies  may  still contain  a few massive black holes  or 
even  satellite galaxies (see below) moving through their extended cores.
Although non-active secondary black holes could be very difficult to   detect directly,
 such systems would be a possible source of jet precession in the AGN of the central galaxy~\citep{RM} or  
 they could induce a detectable  displacement  between the galactic photo-center and its nuclear point source~ \citep{BRAM}.

In the computations presented above  the infalling object was treated as a test particle of fixed mass.
However, in a massive galaxy like M87 the central density is low
enough that the infalling black hole may retain a significant fraction of stars from its host galaxy (because tidal forces are small).
If stalling occurs, then one or more satellites may remain into the core of the central galaxy  for a time significantly longer than a Hubble time.

  \begin{figure}
\begin{center}
 \includegraphics[angle=0,width=2.8in]{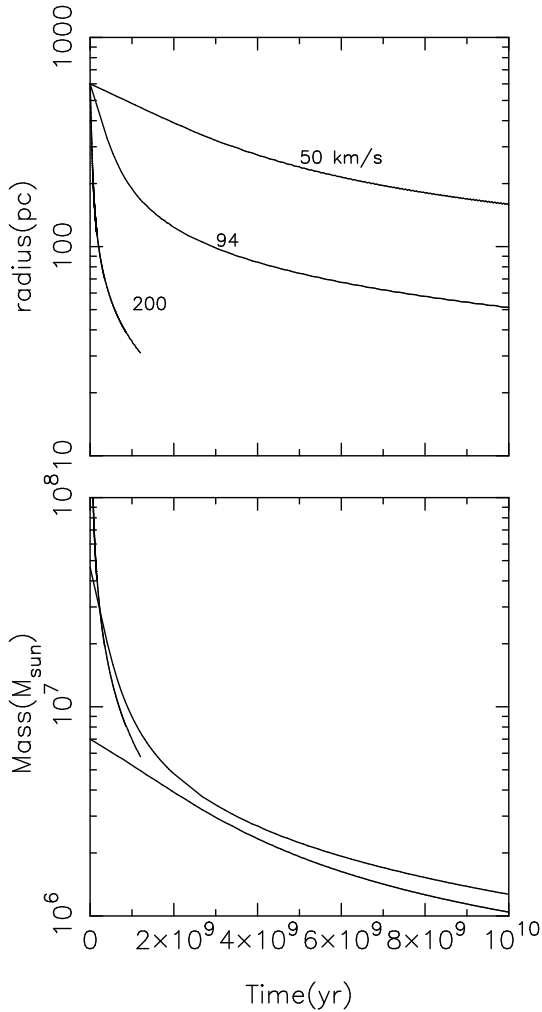} 
   \caption{ Upper panel displays the orbital decay of  satellite galaxies with different  central velocity dispersions (or central black hole masses)
      into the core of  M87. The evolution of the  mass in stars of the infalling galaxies, 
      as determined by the central galaxy tidal field, is given in the lower panel.\\
     }     
  \label{FP}
\end{center}
\end{figure}

To address this possibility we integrated the equations of motion of a satellite galaxy  in a fixed potential 
including the contribution of dynamical friction and the effect of tidal truncation~\citep[e.g.,][]{AA:11}.
The tidally truncated mass of the satellite galaxy~($m_T$) is related to its limiting radius~($r_T$) via:
\begin{equation}
Gm_{T}\approx   \frac12 \sigma^{2}r_{T} \label{eq:mt},
\end{equation}
with $\sigma$ the one dimensional central velocity dispersion. 
The mass of the satellite  SMBH is fixed by 
$\sigma$  through the $M-\sigma$ relation~\citep{GU09}:
\begin{equation}
M = 1.3\times 10^8 (\sigma/200~{\rm km~s^{-1}})^{4.24}.
\end{equation}
The tidal radius can then be related to the potential
$\phi$ and density $\rho$ of the central galaxy by \citep[e.g.,][]{K62}
\begin{equation}
r_{T}=\frac{1}{\sqrt{2}}\sigma\left[\frac{3}{r}\left(\frac{d\phi}{dr}\right)-4\pi G\rho\right]^{-1/2}\label{eq:rt}.
\end{equation}
Using for the central galaxy the mass distribution of equation~(\ref{M87d}),
we find:
\begin{equation}
\frac{d\phi}{dr}=\frac{8\pi}{5}G {\rho_0} {r_0}\left(\frac{r}{{r_0}}\right)^{\frac{1}{2}}+\frac{GM_{\bullet}}{r^{2}}~,
\end{equation}
where $\rho_0=35~M_{\odot}/{\rm pc^{3}}$ and  $r_0=600~$pc.
This gives a limiting radius 
\begin{equation}
r_{T}=\frac{1}{\sqrt{2}}\sigma \left[ \frac{4 \pi}{5} G {\rho_0} \left( \frac{r}{{r_0}} \right)^{-1/2}+
\frac{3GM_{\bullet}}{r^3} \right]^{-1/2}
\end{equation}
and a tidally-truncated mass from equation~(\ref{eq:mt}).
Adopting an initial distance of $r=600~$pc and $\sigma=94~{\rm kms^{-1}}$ (corresponding to $M=4\times10^6~M_{\odot}$) we find  $m_T=4.6\times10^7M_{\odot}$ and  $r_{T}=45~$pc.

Figure~\ref{FP}  plots the orbital evolution of satellites  with initial orbital radius $r=600~$pc and 
different values of the central velocity dispersion $\sigma=(50,~94,~200)~{\rm km~s^{-1}}$ corresponding to 
$M=(3\times10^5,~4\times10^6,~10^8)~M_{\odot}$.
In the core of a giant elliptical galaxy like M87,
the time to reach the center for  galaxies with  $\sigma \lap 100~{\rm km s^{-1}}$ 
is evidently longer than a Hubble time.

 First-ranked galaxies in clusters are often observed to contain multiple ``nuclei,'' which may be identified with the 
 tidally-truncated remains of inspiralling galaxies~\citep{Merritt:84}.
 
 \section{Conclusions} 

In this paper we considered the orbital evolution of massive objects in 
nuclei with shallow density profiles around supermassive black holes (SMBHs).
Our principle results are summarized below.
\begin{itemize}

\item[1] Orbital evolution can be very sensitive to the details of the stellar distribution.
In models with a flat central density profile,  $\rho\sim r^{-\gamma}$,  
$\gamma \approx 0.5$, the dynamical 
friction timescale is much longer than in models with a steep cusp due to
the lack of low-velocity stars.
The standard formula predicts that the inspiraling body will stall
at a radius that is roughly $1/2$ the core radius.

\item[2] Orbital eccentricity increases rapidly when the periapsis 
falls inside the core. 
If the inspiralling body is initially at $r_{\rm bh}$  with  $e_{\rm in} \gtrsim 0.5$, its orbital
eccentricity can become very large ($\gtrsim 0.9$)
 by the time the orbit lies entirely inside the core.

\item[3]  Using $N$-body simulations, we found that the frictional force never falls
precisely to zero.
When the contribution of the fast-moving stars is included in the expression for the dynamical friction force, and (if appropriate) the changes induced by the massive body on the stellar distribution are taken into account,
Chandrasekhar's theory reproduces   the decay observed
in the $N$-body simulations very accurately.
On the other hand, a straightforward application of 
Chandrasekhar's formula in its standard form can give  misleading results.

\item[4]  If the mass of the inspiralling object is sufficiently large, it promotes the diffusion of stars into the phase-space region that was initially nearly empty, increasing
the dynamical friction force. A low-density core is again regenerated as the object
displaces these stars.

\item[5] We derived an estimate of the Coulomb logarithm without any particular assumptions
about the velocity distribution of field stars (e.g., that it follows a Maxwellian distribution), and 
in the region outside the core, where the standard dynamical friction formula (\ref{dfa}) 
accurately represents the motion.
We obtained $\ln \Lambda \approx 6.5$, consistent with previous work.

\item[6] We studied  the location and evolution of the gravitational wake that the
inspiralling body induces in the stellar background.
Outside the core, the peak in the overdensity lies close to the massive body at $-20<\theta \lesssim 0^{\circ}$,  independent of $M$,
and the amplitude of the overdensity increases with black hole mass. After the massive body enters  the core,  the density maximum  
decreases. This is consistent with the fact that the frictional drag is greatly reduced inside the shallow cusp. 

\item[7]  In the absence of a steep central density cusp, the time for 
stellar-mass black holes to reach the center of the Milky Way from a starting 
radius of order $1~\mathrm{pc}$ can easily exceed  $10 {\rm Gyr}$. 
We  computed the evolution  of a population  of stellar black holes 
as they segregate to the Galactic center, including the frictional
force from the fast-moving stars. 
We found that,  in models with parsec-scale cores,  even after $10~$Gyr, the density of black holes can remain  substantially lower than that in stars at all radii.
We conclude that it would be unjustified to assume that the massive remnants have yet reached a steady-state distribution at the Galactic center.

\item[8] Secondary black holes  reach the gravitational radiation dominated regime on orbits that are typically very eccentric.
However, we found that even  initially moderate eccentricities would result in non-negligible eccentricities at the moment the
binary black hole enters the sensitivity window of  planned space-based interferometers. This in turn would require 
non-circular templates for  gravitational wave  data analysis.
\end{itemize} 

As a final remark, we recommend using equation~(\ref{form}) for the study of the inspiral of massive objects in galactic centers.

\bigskip
This work was supported by the National Science Foundation under  
grants no. AST 08-07910 and 08-21141 and by the National Aeronautics and 
Space Administration under grant no. NNX-07AH15G. We thank the referee
M. Weinberg for comments that helped to improve the paper, and we are
indebted to  T.~Alexander,
B.~Kocsis, H.~Perets and E.~Vasiliev for useful discussions.


\end{document}